\documentclass[aps,prc,preprint,groupedaddress]{revtex4}

\usepackage{graphicx}
\usepackage{amsmath}
\usepackage{amsfonts}%
\usepackage{amssymb}
\usepackage{bm}

\begin{document}

\title{$\pi$NN coupling and two-pion photoproduction on the nucleon}

\author{
Michihiro H{\small IRATA}\footnote{hirata@theo.phys.sci.hiroshima-u.ac.jp},
Nobuhiko K{\small ATAGIRI}}

\affiliation{
Department of Physics, Hiroshima University, Higashi-Hiroshima 739, Japan}

\author{
and \\
Takashi T{\small AKAKI}\footnote{takaki@onomichi-u.ac.jp}}
\affiliation{
Department of Economics, Management and Information Science,
Onomichi University, Onomichi 722-8506, Japan}

\date{\today}

\begin{abstract}
Effects of non-resonant photoproductions arising from two different $\pi NN$
couplings are investigated in the $\gamma N\rightarrow\pi\pi N$ reaction. We find
that the pseudoscalar (PS) $\pi NN$ coupling is generally preferable to the
pseudovector (PV) $\pi NN$ coupling and particularly the total cross sections
are successfully described by the model with the PS $\pi NN$ coupling. In
order to see the difference between the two couplings, we also show 
the results of invariant mass spectra and helicity-dependent cross sections
in various isospin channels calculated with the PS and PV couplings.

\end{abstract}


\maketitle

\section{Introduction}

 Many experiments \cite{ABB,pia,gia,car,bra,lan,hae,wolf,zab}of two-pion
photoproductions on the nucleon have been performed in the past over the
second nucleon-resonance energy where the $N^{\ast}(1520)$ resonance plays an
important role. Two charged pion photoproduction, i.e., $\gamma p\rightarrow
\pi^{+}\pi^{-}p$, is well understood by theoretical models. However, none of the
theoretical models have succeeded to explain the data of all isospin channels
simultaneously. This means that there are some unknown production mechanisms
which are not taken into account yet.

Recently, the total cross sections and invariant mass distributions of
two-pion photoproductions followed by neutral pion emission, i.e., $\gamma
p\rightarrow\pi^{+}\pi^{0}n$\cite{lan}, $\gamma p\rightarrow\pi^{0}\pi^{0}%
p$\cite{wolf} and $\gamma n\rightarrow\pi^{-}\pi^{0}p$\cite{zab} reactions,
have been measured at the Mainz accelerator facility MAMI using the detector
system with improved resolution at the photon energy up to around 0.8 GeV. The
aim of these experiments was to obtain the information on the structure of the
nucleon resonance and explore the reaction mechanisms. The $\gamma
p\rightarrow\pi^{+}\pi^{0}n$ and$\quad\gamma n\rightarrow\pi^{-}\pi^{0}p$
reactions have attracted special attention, since the detailed study of them
could provide a new aspect on the reaction mechanisms which are related to
$\rho$ meson production. There are notable differences between the $\pi^{+}%
\pi^{0}$ ($\pi^{-}\pi^{0}$) and $\pi^{0}\pi^{0}$ photoproductions. The $\rho$
meson production as an intermediate process is allowed to the $\pi^{+}\pi^{0}%
$( $\pi^{-}\pi^{0}$) photoproduction but is forbidden to the $\pi^{0}\pi^{0}$
photoproduction due to isospin conservation. On the other hand, the isospin
$I=0$ $\pi\pi$ system such as $\sigma$ meson may contribute to only the
$\pi^{0}\pi^{0}$ photoproduction. Furthermore, the strength of the $\Delta$
Kroll-Rudermann process in the $\pi^{+}\pi^{0}$($\pi^{-}\pi^{0}$)
photoproduction is weak compared with that in the $\pi^{+}\pi^{-}$ photoproduction
where its process dominates and is suppressed in the $\pi^{0}\pi^{0}$
photoproduction. Based on these characteristic features and the comparison
between the measured $\pi^{+}\pi^{0}$ and $\pi^{0}\pi^{0}$
invariant mass distributions, W.Langg\"{a}rtner et al. \cite{lan}concluded that the $\rho$
decay of the $N^{\ast}(1520)$ resonance was directly observed in the $\gamma
p\rightarrow\pi^{+}\pi^{0}n$ reaction.

Theoretically, several works\cite{oset1,laget,ochi1,ochi2,nacher1} have been
already done to explain the total cross sections of two-pion photoproductions
in various isospin channels. The total cross sections of\ the $\gamma
p\rightarrow\pi^{+}\pi^{-}p$ reaction have been well reproduced by several
theoretical models\cite{oset1,laget,ochi1,ochi2,nacher1} where the $\Delta$
Kroll-Rudermann, $\Delta$-pion-pole and $N^{\ast}(1520)$ resonant terms were
found to be dominant. These models, however, could not predict the
total cross sections of the above reactions accompanied by neutral pions
consistently. The models of Tejedor-Oset\cite{oset1} and
Murphy-Laget\cite{laget} underestimated the cross sections of $\gamma
p\rightarrow\pi^{+}\pi^{0}n$ seriously, although the former model could
reproduce the cross sections of $\gamma p\rightarrow\pi^{0}\pi^{0}p$.

The model of Ochi-Hirata-Takaki\cite{ochi1,ochi2} could reproduce the $\gamma
p\rightarrow\pi^{+}\pi^{0}n$ and $\gamma n\rightarrow\pi^{-}\pi^{0}p$ cross
sections as well as $\gamma p\rightarrow\pi^{+}\pi^{-}p$ cross sections,
although it underestimated the $\gamma p\rightarrow\pi^{0}\pi^{0}p$ cross
sections. Their calculations indicated that the $\rho$ meson productions,
i.e., the $N^{\ast}(1520)\rightarrow\rho N$ process and $\rho$
Kroll-Ruderman process, play an essential role in the $\gamma p\rightarrow
\pi^{+}\pi^{0}n$ and $\gamma n\rightarrow\pi^{-}\pi^{0}p$ reactions where
the mass of $\rho$ meson produced in the intermediate state is always smaller than
the on-shell value at the relevant energies. In their model, the $\rho$ meson
is treated in a dynamical way where the finite-range $\rho\pi\pi$ form factor
is assumed. In order to obtain large cross sections of the $\gamma
p\rightarrow\pi^{+}\pi^{0}n$ and $\gamma n\rightarrow\pi^{-}\pi^{0}p$
reactions, a rather soft $\rho\pi\pi$ form factor, which makes the
contribution of the $\rho$ Kroll-Ruderman term large, was needed. Because of
the soft $\rho\pi\pi$ form factor, however, the dynamical model for the $\rho$
meson overestimates the $\pi\pi$ p-wave (isospin $I=1$) phase shifts at low
energies. They speculated that the large $\rho$ Kroll-Ruderman term in their
model might simulate a background process in the isospin $I=1$ channel
effectively rather than the $\rho$ meson production. The presence of such a
background process is inferred from the fact that in the isovector $\pi\pi$
spectral function derived from the $\pi\pi\rightarrow N\overline{N}$ helicity 
amplitudes\cite{toki}, there is a strong enhancement near the $\pi\pi$ threshold
as well as the resonant structure by the $\rho$ meson. This bump at low energies
is actually due to the non-resonant process described by the partial amplitude of
the nucleon Born term projected to the $I=J=1$ $\pi\pi$($N\overline{N}$) channel.

In this paper, motivated by the above speculation, we will investigate the
effect of the non-resonant reaction mechanisms, particularly, the background terms
arising from the $\pi NN$ coupling that were so far considered to have only
small contributions to the cross sections and were always taken to be the
pseudovector (PV) coupling. In the studies of single pion photoproductions,
it has been shown that the PV $\pi NN$ coupling is preferred at low energies
but the pseudoscalar (PS) $\pi NN$ coupling is needed to get a better
description at higher energies\cite{tiator} and furthermore the two $\pi NN$
couplings lead to rather different cross sections for the neutral pion
production\cite{bl}. These facts imply that the PS coupling becomes important
at larger off-shell nucleon momenta and thus will have a significant influence
on the two-pion photoproductions accompanied by the neutral pion such
as the $\gamma p\rightarrow\pi^{+}\pi^{0}n$ and
$\gamma p\rightarrow\pi^{0}\pi^{0}p$ reactions.
In our calculations with two different couplings, we will show that the
non-resonant photoproduction by the PS coupling significantly contributes to
the total cross sections of the two-pion photoproductions involving
the neutral pion
compared with that by the PV coupling
and consequently plays a similar role with the
strong $\rho$ Kroll-Ruderman term introduced in the model by Ochi et
al.\cite{ochi1,ochi2}. From the comparison between the full calculations with
resonant processes and the experimental data, furthermore, we will demonstrate
that the PS coupling is more favored than the PV coupling for the two-pion
photoproductions at the relevant energies.

Recently, Nacher et al.\cite{nacher1} have improved the model of
Tejedor-Oset\cite{oset1} by including the $\Delta(1700)$ production and the
$\rho$ meson effect arising from the $N^{\ast}(1520)$ production. In their
calculations with the PV coupling, they found that the $\rho$ meson effect
largely increased the $\gamma p\rightarrow\pi^{+}\pi^{0}n$ total cross sections
compared with the previous model and put the calculations close to the data,
although there still remained some disagreement with the data around the peak
for both the $\gamma p\rightarrow\pi^{+}\pi^{0}n$ and $\gamma p\rightarrow
\pi^{0}\pi^{0}p$ cross sections. We will discuss why their model can largely
improve the calculations of the $\gamma p\rightarrow\pi^{+}\pi^{0}n$ cross
sections without introducing the strong $\rho$ Kroll-Ruderman term used in the
model by Ochi et al.\cite{ochi1,ochi2}.

In section II, we will discuss the background processes and $\pi NN$ coupling. In
section III, we will review how to treat the resonance processes. In section IV, 
we will show
our full calculations of the total cross sections, invariant mass spectra and
helicity-dependent cross sections and discuss the difference between the PS
and PV couplings from the comparison with the data.\ In section V, we will give our
concluding remarks.

\section{Non-resonant processes and $\pi NN$ coupling\bigskip}

In this section we discuss the non-resonant processes arising from the $\pi
NN$ coupling and vector mesons. There are two types of $\pi NN$ couplings:
the pseudoscalar (PS) coupling and the pseudovector (PV) coupling. The
Lagrangian is%

\begin{equation}
\mathcal{L}_{\pi NN}^{PS}=ig_{\pi NN}\ \overline{\psi}\gamma^{5}\bm{\tau\ }
\psi\bm{\phi}  ,
\end{equation}
for the PS coupling, and
\begin{equation}
\mathcal{L}_{\pi NN}^{PV}=\frac{f_{\pi NN}}{m_{\pi}}\overline{\psi}\gamma
^{\mu}\gamma^{5}\bm{\tau\ }  \psi\partial_{\mu}\bm{\phi}  ,
\end{equation}
for the PV coupling, where $\psi$ and $\bm{\phi}  $ are the nucleon and
pion fields, respectively, $g_{\pi NN}^{2}\ /4\pi=14.4$ and $f_{\pi NN}/m_{\pi
}=-g_{\pi NN}/2M_{N}$ and $m_{\pi}$ and $M_{N}$ are the pion and nucleon
masses, respectively. \ These couplings are equivalent when both the nucleons
are on-shell. If a photon line is attached to the $\pi NN$ system so as to
become gauge-invariant, one obtains the Feynman diagrams of the Born terms for
the one-pion photoproduction whose expressions depend on the $\pi NN$
coupling. The $\pi$ Kroll-Ruderman term is included in the PV Born terms. The
Born terms calculated with the two couplings are rather different each other
because of their different off-shellness. The PV coupling is preferable to the
PS coupling at low energies because the PV Born terms are consistent with low
energy theorems and current algebra predictions. In fact, the model with the PV
coupling is able to reproduce the $E_{0+}$ multipole\ up to the $\Delta(1232)$
resonance energy region but the model with the PS coupling fails. As the
incident photon energy increases over 500 MeV, however, the pure PV coupling
cannot explain the $E_{0+}$ and $M_{1-}$ multipoles and the PS coupling is
needed to describe them\cite{tiator}. This suggests that the $\pi NN$ vertex
for the far off-shell nucleon is largely pseudoscalar in nature. We note that
only the multipoles $E_{0+}$ and $M_{1-}$ are affected by changing the
coupling scheme and the two $\pi NN$ couplings give rise to significantly
different cross sections for the neutral pion photoproduction but are almost
indistinguishable in the charged pion photoproduction. Therefore, it is
interesting to see the difference between the PV and PS couplings in
the two-pion photoproduction around the $N^{\ast}(1520)$ resonance energy
region where the far off-shell nucleons are involved in the intermediate state.

Before going to the $\gamma N\rightarrow\pi\pi N$ reaction, we discuss briefly
the strong interaction part relevant to the two-pion production. In the $\pi
N\rightarrow\pi N$ scattering at low energies, the Born terms constructed with
the PV coupling are also preferable to those with the PS coupling like the
$\gamma N\rightarrow\pi N$ reaction. In this case, for instance, the s-wave
isoscalar scattering length calculated from the PS Born terms is very large
and in disagreement with the data. In dispersion relation theory, the PS Born
terms correspond to nucleon-pole terms and the PV Born terms are understood to
include strong corrections coming from the dispersive integral in addition to
the pole terms. 
These corrections are important to well describe the low energy
data of the $\pi N$ scattering and may be partially related to the $\sigma$
meson exchange contribution as inferred from the linear $\sigma$ model. When
constructing the model for the $\gamma N\rightarrow\pi\pi N$ reaction with the
PS $\pi NN$ coupling, these corrections must be taken into account. To do so,
we introduce the following effective Lagrangian
\begin{equation}
\mathcal{L}_{\pi\pi NN}=\frac{g_{\pi NN}^{2}}{2M_{N}}\overline{\psi}%
\psi\bm{\phi}  \cdot\bm{\phi+}  \frac{g_{\pi NN}^{2}}{(2M_{N})^{2}}%
\overline{\psi}\gamma^{\mu}\bm{\tau\ }  \psi(\bm{\phi}  \times\partial_{\mu
}\bm{\phi}  ). \label{contact}%
\end{equation}
The sum of the PS Born terms and their correction terms calculated from the
above effective Lagrangian are equivalent to the PV Born terms for the strong
interaction processes such as $\pi N\rightarrow\pi N$ and $N\rightarrow\pi\pi
N$.
We note that
the $\pi N$ isoscalar scattering length calculated in these models are
consistent with the value obtained by taking $m_{\sigma}\rightarrow\infty$ in
the linear $\sigma$ model, where $m_{\sigma}$ is the $\sigma$ meson mass.%
\begin{figure}
[ptb]
\begin{center}
\includegraphics[
trim=0.359680in 1.121541in 0.365468in 0.809287in,
height=13.384cm,
width=13.1666cm
]%
{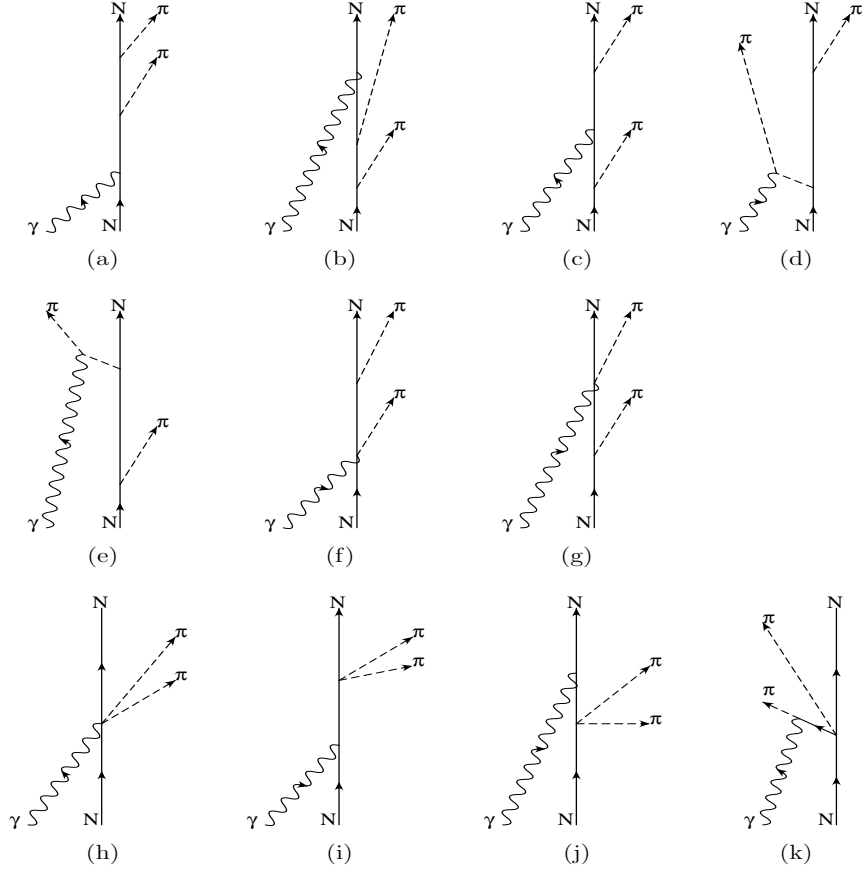}%
\caption{Diagrams of Born terms arising from $\pi NN$ coupling ((a)-(g)) and
the effective Lagrangian of Eq.(\ref{contact}) ((h)-(k)).}%
\label{fig1}%
\end{center}
\end{figure}

The diagrams shown in
Fig.1 for the $\gamma N\rightarrow\pi\pi N$ reaction are obtained by
attaching an external photon line to the diagrams of the $N\rightarrow\pi\pi N$
processes based on the requirement of the gauge invariance. 
The diagrams calculated with the PV coupling correspond to Fig.1(a)-(g)
and especially the diagrams (f) and (g) include the $\pi$ Kroll-Rutherman term
arising from the derivative $\pi NN$ coupling.
On the other hand, the diagrams
calculated with the PS coupling and the above effective Lagrangian correspond
to Fig.1(a)-(e) and (h)-(k), respectively.
They will be referred to as PV model and PS model,
respectively. For the second term (isovector
term) of Eq.(\ref{contact}), only the diagram (k) of Fig.1 is computed in
actual calculations because the contributions of the other diagrams are
negligible. Here the Lagrangians for the $\gamma NN$ and $\gamma\pi\pi$
couplings are given by
\begin{align}
\mathcal{L}_{\gamma NN}  &  =-e\overline{\psi}(F_{1}\gamma^{\mu}A_{\mu}%
-F_{2}\sigma^{\mu\nu}(\partial_{\nu}A_{\mu}))\psi,\label{gnn}\\
\mathcal{L}_{\gamma\pi\pi}  &  =e(\partial_{\mu}\bm{\phi\times\phi)}
_{3}A^{\mu},
\end{align}
where $A_{\mu}$ is the photon field, and $F_{1}$ and $F_{2}$ are the
electromagnetic form factors which are taken to be $F_{1}=1$ and $F_{2}%
=1.79/(2M_{N})$ for the proton and $F_{1}=0$ and $F_{2}=-1.91/(2M_{N})$ for the
neutron, respectively. The other Lagrangians for coupling of a photon and
hadrons are obtained from the minimal substitution ($\partial_{\mu}%
\rightarrow\partial_{\mu}+ieA_{\mu}$) in the Lagrangians of $\pi NN$ and
$\pi\pi NN$ \ derivative vertices. The resultant amplitudes for the $\gamma
N\rightarrow\pi\pi N$ reaction will give rather different cross sections
depending on whether the PS model or PV model is used because of their
different off-shell behavior, as expected from the above discussion about the
$\gamma N\rightarrow\pi N$ reaction. We will show the difference from the
numerical results later.

As other non-resonant processes, we include the contributions of vector mesons
such as $\rho$ meson and $\omega$ meson. In the analysis of the $\gamma N\rightarrow
\pi N$ reaction, these contributions are known to be important although their
magnitude is small. The hadronic Lagrangians involving these vector mesons
are
\begin{align}
\mathcal{L}_{\rho NN}  &  =-\,\overline{\psi}\left(  g_{\rho NN}^{V}%
\ \gamma^{\mu}-\frac{g_{\rho NN}^{T}}{2M_{N}}\sigma^{\mu\nu}\partial_{\nu
}\right)  \bm{\rho}  _{\mu}\cdot\bm{\tau\ }  \psi,\\
\mathcal{L}_{\rho\pi\pi}  &  =-f_{\rho}\bm{\rho}  _{\mu}\cdot(\bm{\phi }
\times\partial^{\mu}\bm{\phi}  ),\label{rho}\\
\mathcal{L}_{\omega NN}  &  =-\,\overline{\psi}\left(  g_{\omega NN}%
^{V}\ \gamma^{\mu}-\frac{g_{\omega NN}^{T}}{2M_{N}}\sigma^{\mu\nu}%
\partial_{\nu}\right)  \omega_{\mu}\bm{\ }  \psi,
\end{align}
where $\bm{\rho}  _{\mu}$ and $\omega_{\mu}$ are the $\rho$ and $\omega$ meson
fields, respectively and the coupling constants are taken to be $g_{\rho
NN}^{V}=2.9,\ g_{\rho NN}^{T}=18.15,\ f_{\rho}=6.0,\ g_{\omega NN}^{V}=7.98$
and $g_{\omega NN}^{T}=0$. The electromagnetic Lagrangian for the $\gamma
\pi_{0}\omega$ coupling is%

\begin{equation}
\mathcal{L}_{\gamma\pi_{0}\omega}=-\,\frac{g_{\gamma\pi\omega}}{m_{\omega}%
}\varepsilon_{\mu\nu\rho\sigma}(\partial^{\mu}A^{\nu})\pi_{0}\partial^{\rho
}(\omega^{\sigma}),
\end{equation}
where $g_{\gamma\pi\omega}=0.374\ e$ and $m_{\omega}=783MeV$. \ The other
electromagnetic Lagrangians are derived from the minimal substitution as done
for the PS and PV models before.
\begin{figure}
[ptb]
\begin{center}
\includegraphics[
trim=0.402676in 4.339977in 0.404330in 0.595270in,
height=10.9172cm,
width=13.3818cm
]%
{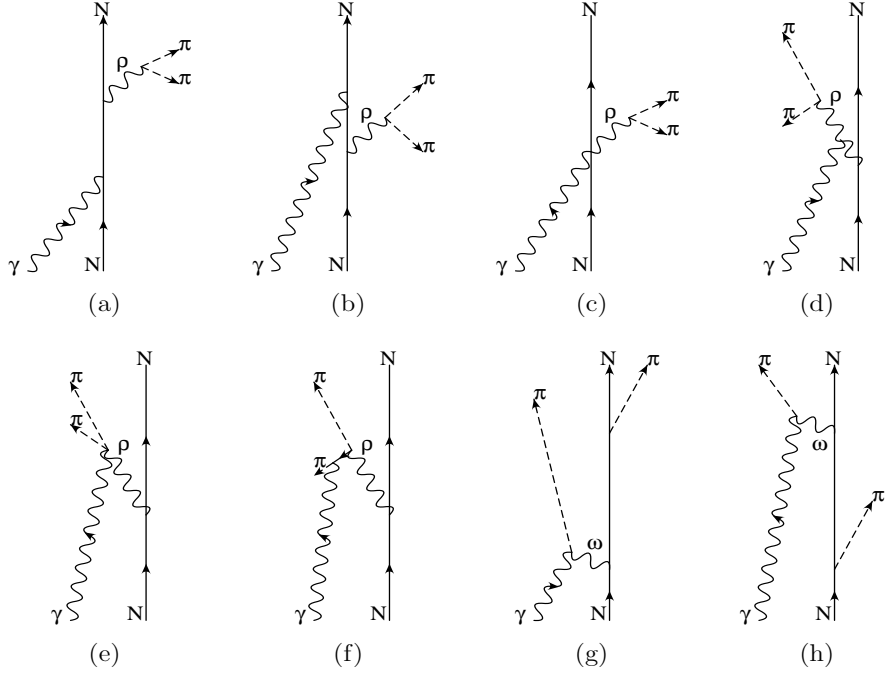}%
\caption{Diagrams of non-resonant processes arising from $\rho$ and $\omega$
mesons.}%
\label{fig2}%
\end{center}
\end{figure}
The diagrams for the $\gamma N\rightarrow\pi\pi N$ reaction involving the
$\rho$ and $\omega$ mesons are shown in Fig.2 and are calculated by using the
above Lagrangians. 
For the $\omega$ meson contribution, diagrams (g) and
(h) in Fig.2 are taken into account. The diagram (c) (called $\rho$
Kroll-Ruderman term) comes from the derivative $\rho NN$ tensor coupling. In
the diagrams of (a)-(d) in Fig.2, the $\rho$ meson decays into two real pions
directly and so the $\rho$ propagator $D_{\rho}$ must include the decay
effect, whose form is assumed to be
\begin{equation}
D_{\rho}(\sqrt{s})=\frac{1}{s-m_{\rho}^{2}+im_{\rho}\,\Gamma_{\rho}(\sqrt{s}%
)},
\end{equation}
with
\begin{equation}
\Gamma_{\rho}(\sqrt{s})=\frac{2}{3}\frac{f_{\rho}^{\ 2}}{4\pi s}%
\mathbf{q}_{cm.}^{3}%
\end{equation}
Here the $\rho$ meson mass $m_{\rho}=775\ $MeV, its width $\Gamma_{\rho
}(m_{\rho})=150$ MeV, and $\mathbf{q}_{cm}$ stands for the pion momentum in
the $\pi\pi$ center of mass system. In our present model, the $\rho$ meson
displayed in Fig.2 is treated in the same way used in Ref.\cite{oset1,nacher1}
where the $\rho\pi\pi$ vertex function is simply given by the Lagrangian of
Eq.(\ref{rho}). \ Even though one uses the $\rho\pi\pi$ vertex function with
the finite-range form factor employed in Ref.\cite{ochi1,ochi2}, there is no
drastic numerical change from the above way in the magnitude of the $\rho$
Kroll-Ruderman term as far as one uses the range parameter determined so as to
reproduce the $\pi\pi$ p-wave (isospin $I=1$) phase shifts.

For the diagrams involving the off-shell meson coupled to nucleon in Fig.1 and
Fig.2, \ we take into account a form factor of the following form:
\begin{equation}
F(q^{2})=\frac{\Lambda^{2}-m^{2}}{\Lambda^{2}-q^{2}},
\end{equation}
where $m$ and $q^{2}$ are the meson mass and the square of four-momentum,
respectively and the range parameter $\Lambda$ is taken to be $\Lambda=1.25$
GeV for $\pi$, $\Lambda=1.4$ GeV for both $\rho$ and $\omega$,
respectively\cite{nacher1}. The pion form factor is used for the diagrams (d)
and (e) in Fig.1 and the $\rho$ meson and $\omega$ meson form factors are used
for (a)-(f) and (g)-(h) in Fig.2, respectively. For the diagrams (f) and (g)
in Fig.1 ($\pi$ Kroll-Ruderman term), the same pion form factor is used and
evaluated at the momentum transfer between the incident photon and the
outgoing pion. The diagram (c) in Fig.2 ($\rho$ Kroll-Ruderman term) is
treated in the same way as the $\pi$ Kroll-Ruderman term. We note that the
gauge invariance of the transition amplitudes is destroyed by the inclusion of
the decay width in the propagator and the form factor for the hadronic vertex.
However, we consider such strong interaction corrections are more important
than the gauge invariance from a phenomenological point of view. In fact, the
form factor used in the $\Delta$ Kroll-Ruderman term (see Fig.5(a)) influences
significantly on the magnitude of the $\gamma p\rightarrow\pi^{+}\pi^{-}p$
cross section and provides a good agreement with the data\cite{oset1}.

Now we discuss the differences between the PS and PV models through numerical
calculations of the cross sections. The cross section for the $\gamma
N\rightarrow\pi_{\alpha}\pi_{\beta}N$ reaction is given by
\begin{align}
\sigma &  =\frac{1}{2|\mathbf{k}|}\frac{M_{N}}{E_{p_{1}}}\frac{1}{v_{rel}}%
\int\frac{d^{3}p_{2}}{(2\pi)^{3}}\frac{d^{3}q_{\alpha}}{(2\pi)^{3}}%
\frac{d^{3}q_{\beta}}{(2\pi)^{3}}\frac{M_{N}}{E_{p_{2}}}\frac{1}%
{2\omega_{\alpha}}\frac{1}{2\omega_{\beta}}\nonumber\\
&  \times(2\pi)^{4}\delta^{(4)}(p_{1}+k-p_{2}-q_{\alpha}-q_{\beta})\sum
_{\nu\nu^{\prime}}\frac{1}{2}|\ \langle1/2,\nu^{\prime}|\ T\ |1/2,\nu
\rangle\ |^{2},
\end{align}
where $k=(|\mathbf{k}|,\mathbf{k})$, $p_{1}=(E_{p_{1}},\mathbf{p}_{1})$,
$p_{2}=(E_{p_{2}},\mathbf{p}_{2})$, and $q_{\gamma}=(\omega_{\gamma
},\mathbf{q}_{\gamma})\ (\gamma=\alpha,\beta)$ are the four-momenta of\ the
initial photon, initial nucleon, final nucleon, and outgoing pion,
respectively, $v_{rel}$ is the relative velocity between the initial nucleon
and photon, and $\omega_{\gamma}=\sqrt{m_{\pi}^{2}+\mathbf{q}_{\gamma}^{2}}$
and $E_{p}=\sqrt{M_{N}^{2}+p^{2}}$. The cross section is evaluated in the
$\gamma N$ center-of-mass system. The T matrix is in general expressed as
$T=A+i\bm{\sigma}  \cdot\mathbf{B}$, which is summed over the final nucleon
spin states $(\nu^{\prime})$ and averaged over the initial nucleon spin states
$(\nu)$.

We show how to evaluate the T matrix by taking one of the diagrams as an
example. Let us consider the process corresponding to (a) in Fig.1 computed
with the PS coupling. The T matrix is divided into two parts of the hadronic
process\ and electromagnetic process and then into the particle and
anti-particle intermediate states for the convenience. Thus, the T matrix is
expressed as
\begin{align}
\chi^{\dagger}T\chi &  =H_{\gamma NN}^{(1)}\frac{1}{E_{p_{2}}-k-E_{p_{2}-k}%
}\,\frac{M_{N}}{E_{p_{2}-k}}\,H_{NN\pi\pi}^{(1)}\nonumber\\
&  +H_{\gamma NN}^{(2)}\frac{1}{E_{p_{2}}-k+E_{p_{2}-k}}\,\frac{M_{N}%
}{E_{p_{2}-k}}\,H_{NN\pi\pi}^{(2)},
\end{align}
where the transition matrices $H$ are given by
\begin{align}
H_{\gamma NN}^{(1)}  &  =e\overline{u}(\mathbf{p}_{2})(F_{1}\bm{\gamma }
\cdot\bm{\varepsilon}  -iF_{2}\sigma^{i\nu}\varepsilon_{i}k_{\nu}%
)u(\mathbf{p}_{2}-\mathbf{k})\nonumber\\
&  \simeq\chi^{\dagger}e(S_{E}^{(1)}+i\bm{\sigma}  \cdot\mathbf{V}_{E}%
^{(1)})\chi,
\end{align}
with%
\begin{align}
S_{E}^{(1)}  &  =\sqrt{\tfrac{E_{p_{2}}+M_{N}}{2M_{N}}}\sqrt{\tfrac
{E_{p_{2}-k}+M_{N}}{2M_{N}}}\left(  \frac{F_{1}+F_{2}k}{E_{p_{2}-k}+M_{N}%
}+\frac{F_{1}-F_{2}k}{E_{p_{2}}+M_{N}}\right)  \mathbf{p}_{2}\cdot
\bm{\varepsilon,} \\
\mathbf{V}_{E}^{(1)}  &  =\sqrt{\tfrac{E_{p_{2}}+M_{N}}{2M_{N}}}\sqrt
{\tfrac{E_{p_{2}-k}+M_{N}}{2M_{N}}}{\Huge [}\frac{F_{1}+(E_{p_{2}-k}%
+M_{N}+k)F_{2}}{E_{p_{2}-k}+M_{N}}\mathbf{k}\times\bm{\varepsilon }
\nonumber\\
&  -\left(  \frac{F_{1}+F_{2}k}{E_{p_{2}-k}+M_{N}}+\frac{F_{1}-F_{2}%
k}{E_{p_{2}}+M_{N}}\right)  \mathbf{p}_{2}\times\bm{\varepsilon}{\Huge ]}  ,
\end{align}
for the $\gamma N\rightarrow N$ process,
\begin{align}
H_{NN\pi\pi}^{(1)}  &  =\overline{u}(\mathbf{p}_{2}-\mathbf{k})g_{\pi
NN}\gamma^{5}\tau_{\beta}\,\frac{/\!\!\!p_{1}-/\!\!\!q_{\alpha}+M_{N}}%
{(p_{1}-q_{\alpha})^{2}-M_{N}^{2}}\,g_{\pi NN}\gamma^{5}\tau_{\alpha
}u(\mathbf{p}_{1})\nonumber\\
&  \simeq\chi^{\dagger}g_{\pi NN}^{2}(S_{1}^{(1)}+i\bm{\sigma}  \cdot
\mathbf{V}_{1}^{(1)})\chi,
\end{align}
with
\begin{align}
S_{1}^{(1)}  &  =\sqrt{\tfrac{E_{p_{2}-k}+M_{N}}{2M_{N}}}\sqrt{\tfrac
{E_{p_{1}}+M_{N}}{2M_{N}}{\Huge [}}\omega_{\alpha}-\left(  \frac{\mathbf{q}%
_{\alpha}\cdot\mathbf{p}_{1}}{E_{p_{1}}+M_{N}}+\frac{\mathbf{q}_{\alpha}%
\cdot(\mathbf{p}_{1}-\mathbf{k})}{E_{p_{1}-k}+M_{N}}\right)  {\Huge ]}%
\nonumber\\
&  \times\frac{\tau_{\beta}\tau_{\alpha}}{(p_{1}-q_{\alpha})^{2}-M_{N}^{2}},\\
\mathbf{V}_{1}^{(1)}  &  =\sqrt{\tfrac{E_{p_{2}-k}+M_{N}}{2M_{N}}}\sqrt
{\tfrac{E_{p_{1}}+M_{N}}{2M_{N}}}\,{\Huge [}\left(  \frac{1}{E_{p_{2}-k}%
+M}-\frac{1}{E_{p_{1}}+M}\right)  \,(\mathbf{q}_{\alpha}\times\mathbf{p}%
_{1})\nonumber\\
&  -\frac{\mathbf{q}_{\alpha}\times\mathbf{q}_{\beta}}{E_{p_{2}-k}+M_{N}%
}{\Huge ]}\,\frac{\tau_{\beta}\tau_{\alpha}}{(p_{1}-q_{\alpha})^{2}-M_{N}^{2}%
},
\end{align}
for the $N\rightarrow\pi\pi N$ process,
\begin{align}
H_{\gamma NN}^{(2)}  &  =\overline{u}(\mathbf{p}_{2})e(F_{1}\bm{\gamma }
\cdot\bm{\varepsilon}  -iF_{2}\sigma^{i\nu}\varepsilon_{i}k_{\nu
})v(-\mathbf{p}_{2}+\mathbf{k})\nonumber\\
&  \simeq\chi^{\dagger}e(-i)(S_{E}^{(2)}+i\bm{\sigma}  \cdot\mathbf{V}%
_{E}^{(2)})\chi,
\end{align}
with%
\begin{align}
S_{E}^{(2)}  &  =\sqrt{\tfrac{E_{p_{2}}+M_{N}}{2M_{N}}}\sqrt{\tfrac
{E_{p_{2}-k}+M_{N}}{2M_{N}}}\left(  \frac{F_{2}}{E_{p_{2}-k}+M_{N}%
}+\frac{F_{2}}{E_{p_{2}}+M_{N}}\right)  (\mathbf{k}\times\bm{\varepsilon }
)\cdot\mathbf{p}_{2},\\
\mathbf{V}_{E}^{(2)}  &  =\sqrt{\tfrac{E_{p_{2}}+M_{N}}{2M_{N}}}\sqrt
{\tfrac{E_{p_{2}-k}+M_{N}}{2M_{N}}}{\Huge [}\left\{  F_{1}+\left(
k-\frac{k^{2}}{E_{p_{2}-k}+M_{N}}\right)  F_{2}\right\}  \cdot\bm{\varepsilon}
\nonumber\\
&  +F_{2}\left(  \frac{1}{E_{p_{2}-k}+M_{N}}-\frac{1}{E_{p_{2}}+M_{N}}\right)
(\mathbf{k}\times\bm{\varepsilon}  )\times\mathbf{p}_{2}{\Huge ]},
\end{align}
for the $\gamma\rightarrow N\bar{N}$ process and
\begin{align}
H_{NN\pi\pi}^{(2)}  &  =\overline{v}(-\mathbf{p}_{2}+\mathbf{k})g_{\pi
NN}\gamma^{5}\tau_{\beta}i\frac{/\!\!\!p_{1}-/\!\!\!q_{\alpha}+M}%
{(p_{1}-q_{\alpha})^{2}-M^{2}}\,g_{\pi NN}\gamma^{5}\tau_{\alpha}%
u(\mathbf{p}_{1})\nonumber\\
&  \simeq\chi^{\dagger}g_{\pi NN}^{2}(-i)(S_{1}^{(2)}+i\bm{\sigma}
\cdot\mathbf{V}_{1}^{(2)})\chi,
\end{align}
with%
\begin{align}
S_{1}^{(2)}  &  =0,\\
\mathbf{V}_{1}^{(2)}  &  =-\sqrt{\tfrac{E_{p_{2}-k}+M_{N}}{2M_{N}}}%
\sqrt{\tfrac{E_{p_{1}}+M_{N}}{2M_{N}}}{\Huge [}\mathbf{q}_{\alpha}+\left(
\frac{1}{E_{p_{2}-k}+M_{N}}-\frac{1}{E_{p_{1}}+M_{N}}\right)  \omega_{\alpha
}\mathbf{p}_{1}\nonumber\\
&  -\frac{\omega_{\alpha}(\mathbf{q}_{\alpha}+\mathbf{q}_{\beta})}{E_{p_{2}%
-k}+M_{N}}{\Huge ]}\frac{\tau_{\beta}\tau_{\alpha}}{(p_{1}-q_{\alpha}%
)^{2}-M_{N}^{2}},
\end{align}
for the $N\bar{N}$ $\rightarrow\pi\pi$ process, respectively. Here $u$ and $v$
are the Dirac spinors for nucleon and anti-nucleon, respectively and$\ \chi$ is
the two-component spinor. In this expression, we neglect the $O((E_{p}%
+M_{N})^{-2})$ contributions. However, we found this approximate T matrix
gives nearly the same result as the exact one within the present energy
region. The calculations for the other diagrams are performed in a similar
fashion.\ In this sense, the T matrix is evaluated in a relativistic way.

In order to estimate the relativistic effect, we calculated the cross section
with the PV coupling for the diagrams (a)-(g) in Fig.1 in both relativistic and
non-relativistic ways. Here non-relativistic approximations mean that both the
anti-particle contributions and terms of order $(p/M_{N})^{2}$ or higher in
vertex operators are neglected but the denominators of the propagator are
treated in a relativistic way. We found that the relativistic calculations are
about 30-60$\%$ larger than the non-relativistic calculations around the
$N^{\ast}(1520)$ resonance energy. This indicates that the relativistic
effect, which have been so far neglected in previous studies, should not be
discarded for the non-resonant process in the $\gamma N\rightarrow\pi\pi N$ reaction.

\begin{figure}[ptb]
\begin{center}
\scalebox{0.6}{\includegraphics{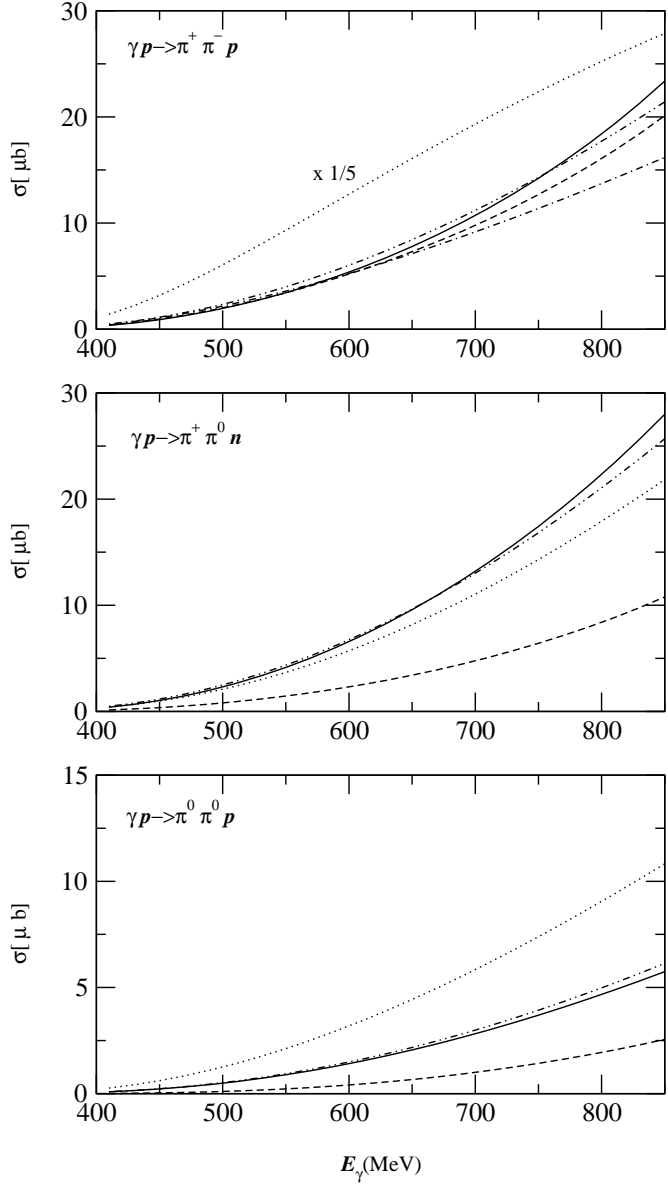}}
\end{center}
\caption{Non-resonant contributions to total cross sections for $\gamma
p\rightarrow\pi^{+}\pi^{-}p$, $\gamma p\rightarrow\pi^{+}\pi^{0}n$ and
.$\gamma p\rightarrow\pi^{0}\pi^{0}p$ reactions. The solid and dashed lines
correspond to the PS and PV calculations with the vector meson contributions,
respectively. The meaning of remaining lines is given in the text.}%
\label{fig3}%
\end{figure}

Now we show the calculations of total cross sections for three isospin
channels: $\gamma p\rightarrow\pi^{+}\pi^{-}p$, $\gamma p\rightarrow\pi^{+}%
\pi^{0}n$ and $\gamma p\rightarrow\pi^{0}\pi^{0}p$ in Fig.3. Here solid and dashed
lines
correspond to the full non-resonant calculations for the PS and PV
models including the vector meson contributions (Fig.2), respectively.
We observe that the PV results of the $\pi^{+}\pi^{-}$
channel are close to the PS results, while in the $\pi
^{+}\pi^{0}$ and $\pi^{0}\pi^{0}$ channels there are significant differences
and the PS calculations are larger than the PV calculations. The similar
feature can be seen in the one-pion photoproduction where the difference
between two couplings is prominent in the neutral pion photoproduction but is
very small in the charged pion photoproduction. This arises certainly from the
different off-shell behavior between the PV and PS $\pi NN$ couplings. For the
one-pion photoproduction, such difference appears in the Born terms
proportional to the anomalous magnetic moment obtained using the second term
of Eq.(\ref{gnn}). In fact, if $F_{2}$ is set to zero, the difference between
the PS and PV calculations disappears. The same things happen for the two-pion
photoproduction, which has been examined numerically.

It is interesting to observe that the PS calculation in the $\gamma
p\rightarrow\pi^{+}\pi^{0}n$ channel is remarkably larger than the PV
calculation and is roughly consistent with the size of the $\rho$
Kroll-Ruderman term introduced in the model of Ochi et al.\cite{ochi1,ochi2},
by which the large measured cross sections have been successfully explained.
In order to reproduce the data, however, the small range-paramater of
the $\rho\pi\pi$
form factor must have been used and it gave rise to the larger $\pi\pi$ scattering
p-wave phase shift at low energies compared with the experimental value. We
think that the $\rho$ Kroll-Ruderman term with this form factor simulates
a non-negligible background process originating from the strong coupling
between the nucleon and $I=J=1$
$\pi\pi$ system as is inferred from strong enhancement at low energies in the
isovector $\pi\pi$ spectral function derived from the $\pi\pi\rightarrow
N\overline{N}$ helicity amplitude\cite{toki}. Therefore, the $\rho$
Kroll-Ruderman term is considered to represent the non-resonant process
arising from the PS coupling effectively.

In Fig.3, we also show the results for the Born terms coming from the PS
coupling (dotted lines) corresponding to the diagrams (a)-(e) in Fig.1. We
find that the contribution to the $\gamma p\rightarrow\pi^{+}\pi^{-}p$ channel
is extremely large and is mostly attributed to the pion pole terms (diagrams
(d) and (e) in Fig.1). The similar situation occurs when one calculates the
$\pi N$ isoscalar s-wave scattering length with the PS coupling. These
unfavorable results can be improved by introducing the contact interaction of
Eq.(\ref{contact}) as mentioned before. This effect is seen in the
calculations with the contact terms (dash-two-dotted lines in Fig.3). Here the
contact terms correspond to the diagrams (h)-(k) in Fig.1 and consist of the
isoscalar and isovector parts. The isoscalar term contributes to the $\gamma
p\rightarrow\pi^{+}\pi^{-}p$ and $\gamma p\rightarrow\pi^{0}\pi^{0}n$ channels
and the isovector term contributes to the $\gamma p\rightarrow\pi^{+}\pi^{-}p$
and $\gamma p\rightarrow\pi^{+}\pi^{0}n$ channels, and the isoscalar term has
larger coupling constant than the isovector term as understood from
Eq.(\ref{contact}). The size of these terms can be seen in the calculations in Fig.3.

For comparison, the result for the Born terms coming from the PV coupling
(diagrams (a)-(g) in Fig.1) is also shown by dash-dotted line in Fig.3. The
dash-dotted line is drawn only in the $\gamma p\rightarrow\pi^{+}\pi^{-}p$
channel but omitted in other channels since the full non-resonant calculations
(dashed line) in other channels are almost overlapped with the dash-dotted
line. The difference between the solid and dash-two-dotted lines or between
the dashed and dash-dotted lines arises from the vector meson contributions of
the diagrams in Fig.2. The contributions are relatively small. This smallness
is mainly attributed to the effect of form factors used. \ In this paper, we
do not pursue this effect furthermore.%
\begin{figure}
[ptb]
\begin{center}
\includegraphics[
height=2.7025in,
width=5.3229in
]%
{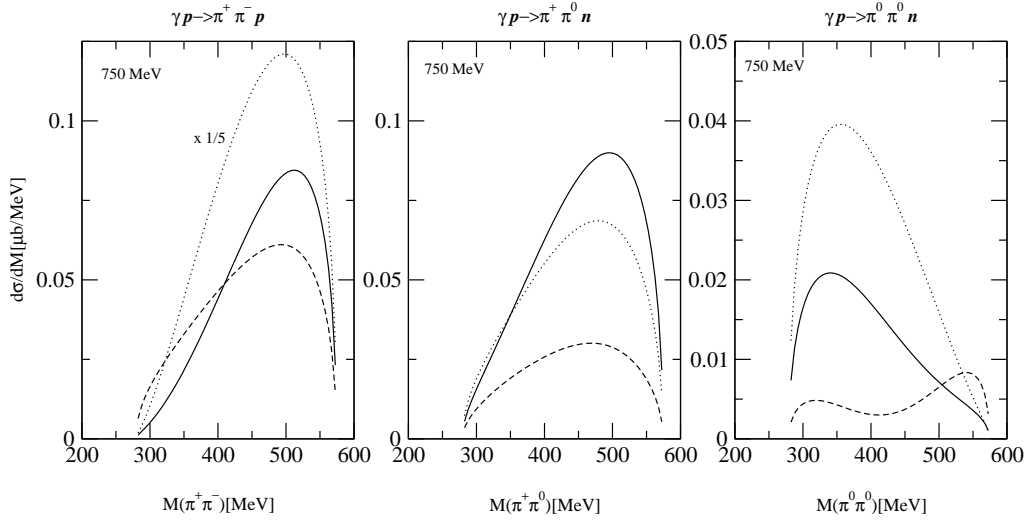}%
\caption{Non-resonant contributions to the $\pi\pi$ invariant mass spectra for
$\gamma p\rightarrow\pi^{+}\pi^{-}p$, $\gamma p\rightarrow\pi^{+}\pi^{0}n$ and
$\gamma p\rightarrow\pi^{0}\pi^{0}p$ reactions at 750 MeV. The meaning of
lines is the same as that of Fig.3.}%
\label{fig4}%
\end{center}
\end{figure}

In order to see the difference between the PS and PV models in further
detail, \ the $\pi\pi$ invariant-mass spectra are calculated and the results
at $750\ $MeV are shown in Fig. 4. The meaning of lines is the same as that of
Fig.3. Apart from the size of the distributions, \ the difference of the
shape can be seen in these distributions, especially in the $\gamma
p\rightarrow\pi^{0}\pi^{0}p$ channel.\ Even in the PS calculations for the
$\gamma p\rightarrow\pi^{+}\pi^{-}p$ and $\gamma p\rightarrow\pi^{+}\pi^{0}n$
channels, there are some shifts of the peak to the higher invariant mass
compared with the PV calculations. Here it is interesting to note that the
$I=J=1\ \pi\pi$ system relevant to the $\rho$ meson can contribute to the
$\gamma p\rightarrow\pi^{+}\pi^{-}p$ and $\gamma p\rightarrow\pi^{+}\pi^{0}n$
reactions and on the other hand the $I=J=0\ \pi\pi$ system relevant to the
$\sigma$ meson can contribute to the $\gamma p\rightarrow\pi^{+}\pi^{-}p$ and
$\gamma p\rightarrow\pi^{0}\pi^{0}p$ reactions. The correlations for the final
$\pi\pi$ system might influence both the shape and size of the distributions
but in our calculations they are not taken into account. In this work, as a
first step, we would like to demonstrate how different the PS model and PV
model are within our present framework. In order to compare the calculations with 
the experimental data, the resonant contributions must be included.
We will employ a simple model
for the resonant processes, which will be discussed in next section.

\section{Resonant processes}

\subsection{isobar model}

The two-pion photoproduction in the second resonance energy region involves
the resonances such as $\rho$ meson, $\Delta(1232)$ and $N^{\ast}(1520)$ as
important intermediate states. We treat these resonances with the isobar model
where the scattering of $\pi\pi$ or $\pi N$ in the relevant channel is assumed
to be described solely by the resonant state. We employ the same formalism
used in Ref.\cite{ochi1,ochi2}, which is briefly reviewed in this section.

In this model, the $\pi\pi$ p-wave scattering $t$ matrix in the energies from
threshold to the $\rho$ meson resonance is assumed to be written as
\begin{equation}
t_{\pi\pi}=\frac{F_{\rho\pi\pi}F_{\rho\pi\pi}^{\dag}}{2m_{\rho}^{0}(\sqrt
{s}-m_{\rho}^{0}-\Sigma_{\rho\pi\pi})}, \label{rho-isobar}%
\end{equation}
where $m_{\rho}^{0}$ and $\sqrt{s}$ denote the bare mass of $\rho$ meson and
the $\pi\pi$ center-of-mass energy, respectively. The $\rho\pi\pi$ vertex
function is assumed to have the form%

\begin{align}
F_{\rho\pi\pi}  &  =2h_{\rho}(\kappa)(\bm{\varepsilon}  _{\rho}\cdot
\bm{\kappa}  ),\nonumber\\
h_{\rho}(\kappa)  &  =\frac{f_{\rho\pi\pi}}{1+(\kappa/q_{\rho\pi\pi})^{2}},
\end{align}
where $f_{\rho\pi\pi}$ and $q_{\rho\pi\pi}$ are the $\rho\pi\pi$ coupling
constant and range parameter, respectively and $\bm{\varepsilon}  _{\rho}$ is
the polarization vector and $\bm{\kappa}  $ is the relative momentum between
two pions. The $\rho$ self-energy $\Sigma_{\rho\pi\pi}$ is evaluated with the
same way used in Ref.\cite{ochi1}. The parameters $m_{\rho}^{0}$,
$\ f_{\rho\pi\pi}$and $q_{\rho\pi\pi}$ are determined to fit the $\pi\pi$
p-wave phase shifts as well as the mass and width of $\rho$ meson. We take
$m_{\rho}^{0}=910$ MeV, $\ f_{\rho\pi\pi}=7.8$ and $q_{\rho\pi\pi}=800$ MeV/c,
which are used in the calculation of the $N^{\ast}$ self-energy. In the
previous paper\cite{ochi1,ochi2}, $q_{\rho\pi\pi}$ was adjusted to fit the
$\gamma p\rightarrow\pi^{+}\pi^{0}n$ data instead of the $\pi\pi$ p-wave phase
shifts. In the case of Ref.\cite{ochi2}, $q_{\rho\pi\pi}$ was taken to be 200
MeV/c, which made the size of the $\rho$ Kroll-Ruderman term quite large. This
parameterization was only a way to phenomenologically reproduce the $\gamma
p\rightarrow\pi^{+}\pi^{0}n$ reaction cross section within the previous model.

The $\pi N$ scattering $t$ matrix in the P$_{33}$channel is written as
\begin{equation}
t_{P33}=\frac{F_{\pi N\Delta}F_{\pi N\Delta}^{\dag}}{\sqrt{s}-M_{\Delta}%
^{0}-\Sigma_{\pi N}^{\Delta}}, \label{delta}%
\end{equation}
where $\sqrt{s}$ and $M_{\Delta}^{0}$ denote the $\pi N$ center-of-mass energy
and bare mass of $\Delta(1232)$, respectively. The vertex function for the
$\pi N\rightarrow$ $\Delta$ transition is expressed as
\begin{equation}
F_{\pi N\Delta}^{\dag}=-i\sqrt{6\pi^{2}}\sqrt{\frac{2\omega_{\pi}E_{p}}{M_{N}%
}}g_{\pi N\Delta}(p)(\mathbf{S}^{\dagger}\cdot\mathbf{\hat{p}),}%
\end{equation}
where $p$ is the three-momentum in the $\pi N$ center of mass system and
$\mathbf{\hat{p}}$ is its unit vector. $\mathbf{S}^{\dagger}$ is the spin
transition operator from 1/2 to 3/2 and $g_{\pi N\Delta}$ is given
by\cite{betz}
\begin{equation}
g_{\pi N\Delta}(p)=\frac{F_{\Delta}}{\sqrt{2(m_{\pi}+M_{N})}}\frac{p}{m_{\pi}%
}\left(  \frac{Q_{\Delta}^{2}}{Q_{\Delta}^{2}+p^{2}}\right)  ^{2},
\label{qdelta}%
\end{equation}
where $F_{\Delta}$ is the coupling constant and $Q_{\Delta}$ is the range
parameter. The $\Delta$ self-energy $\Sigma_{\pi N}^{\Delta}$ is evaluated
using the vertex function $F_{\pi N\Delta}^{\dag}$. The parameters $M_{\Delta
}^{0}$, $F_{\Delta}$ and $Q_{\Delta}$ were adjusted to fit the experimental
P$_{33\text{ }}$scattering amplitude\cite{betz}.

The $\pi N$ scattering $t$ matrix in the D$_{13}$ channel has the same form
with the the above P$_{33}$ amplitude $t_{P33\text{ }}$as follows:%

\begin{equation}
t_{D13}=\frac{F_{\pi NN^{\ast}}F_{\pi NN^{\ast}}^{\dag}}{\sqrt{s}-M_{N^{\ast}%
}^{0}-\Sigma_{\mathrm{total}}^{N^{\ast}}},
\end{equation}
where $M_{N^{\ast}}^{0}$ denotes the bare mass of $N^{\ast}$. The vertex
function for the $\pi N\rightarrow N^{\ast}$ transition is written
as\cite{arima}
\begin{equation}
F_{\pi NN^{\ast}}^{\dag}=-i(2\pi)^{3/2}\sqrt{\frac{2\omega_{\pi}E_{p}}{M_{N}}%
}\frac{f_{\pi NN^{\ast}}}{\sqrt{2(m_{\pi}+M_{N})}}\left(  \frac{p}{p_{\pi
NN^{\ast}}}\right)  ^{2}e^{-(p/p_{\pi NN^{\ast}})^{2}}\left(  S^{(2)\dag}\cdot
Y_{2}(\widehat{\mathbf{p}})\right)  ,
\end{equation}
where $f_{\pi NN^{\ast}}$ is the $\pi NN^{\ast}$ coupling constant and $p_{\pi
NN^{\ast}}$ is the $\pi NN^{\ast}$ range parameter. $S^{(2)\dag}$ is the
second-rank spin transition operator from 1/2 to 3/2, which is defined by%

\begin{equation}
S^{(2)\dag}=\sqrt{\frac{2}{5}}\left[  \mathbf{S}^{\dag}\times\bm{\sigma }
\right]  ^{(2)}.
\end{equation}
The $N^{\ast}(1520)$ resonance can decay into both the $\pi N$ and $\pi\pi N$
channels. The $\pi\pi N$ decay occurs through three dominant modes, i.e.,
$(\pi\Delta)_{s-wave}$, $(\pi\Delta)_{d-wave}$ and $\rho N$ . These branching
fractions are known to be comparable. Thus the total $N^{\ast}$ self-energy
($\Sigma_{\mathrm{total}}$) is expressed as
\begin{equation}
\Sigma_{\mathrm{total}}^{N^{\ast}}=\Sigma_{\pi N}^{N^{\ast}}+\Sigma_{\pi
\Delta}^{s}+\Sigma_{\pi\Delta}^{d}+\Sigma_{\rho N},
\end{equation}
where $\Sigma_{\pi N}^{N^{\ast}}$, $\Sigma_{\pi\Delta}^{s}$, $\Sigma
_{\pi\Delta}^{d}$ and $\Sigma_{\rho N}$ are due to the coupling to the $\pi
N$, s-wave $\pi\Delta$, d-wave $\pi\Delta$ and $\rho N$ channels,
respectively. $\Sigma_{\pi N}^{N^{\ast}}$ is evaluated from the vertex
function $F_{\pi NN^{\ast}}^{\dag}$ and the other components of the
self-energy are obtained from the following vertex functions:%

\begin{equation}
F_{\pi\Delta N^{\ast}}^{s\dag}(p)=-i(2\pi)^{3/2}\sqrt{\frac{2\omega_{\pi}%
}{2(m_{\pi}+M_{N})}}f_{\pi\Delta N^{\ast}}^{s}e^{-\left(  p/p_{\pi\Delta
N^{\ast}}^{s}\right)  ^{2}}Y_{00}(\widehat{\mathbf{p}}),
\end{equation}
for the $N^{\ast}\rightarrow(\pi\Delta)_{s-wave}$,%

\begin{equation}
F_{\pi\Delta N^{\ast}}^{d\dag}(p)=-i(2\pi)^{3/2}\sqrt{\frac{2\omega_{\pi}%
}{2(m_{\pi}+M_{N})}}f_{\pi\Delta N^{\ast}}^{d}\left(  \frac{p}{p_{\pi\Delta
N^{\ast}}^{d}}\right)  ^{2}e^{-(p/p_{\pi\Delta N^{\ast}}^{d})^{2}}\left(
S_{3/2}^{(2)\dag}\cdot Y_{2}(\widehat{\mathbf{p}})\right)  ,
\end{equation}
for the $N^{\ast}\rightarrow(\pi\Delta)_{d-wave}$ and
\begin{equation}
F_{\rho NN^{\ast}}^{\dag}=(2\pi)^{3/2}\sqrt{\frac{2\omega_{\rho}E_{p}}{M_{N}}%
}f_{\rho NN^{\ast}}e^{-(p/p_{\rho NN^{\ast}})^{2}}\left(  \mathbf{S}{^{\dag}%
}\cdot\bm{\varepsilon}{_{\rho}}  \right)  Y_{00}(\widehat{\mathbf{p}}),
\end{equation}
for the $N^{\ast}\rightarrow\rho N$ . Here $f_{\pi\Delta N^{\ast}}^{s}$ ,
$f_{\pi\Delta N^{\ast}}^{d}$ and $f_{\rho NN^{\ast}}$ are the s-wave, d-wave
$\pi\Delta N^{\ast}$ and $\rho NN^{\ast}$ coupling constants, respectively.
$p_{\pi\Delta N^{\ast}}^{s,d}$ and $p_{\rho NN^{\ast}}$ are the $\pi\Delta
N^{\ast}$ and $\rho NN^{\ast}$ range parameters. $S_{3/2}^{(2)\dag}$ is the
second-rank spin transition operator from 3/2 to 3/2 defined in
Ref.\cite{ochi1}. $\Sigma_{\pi\Delta}^{s(d)}$ and $\Sigma_{\rho N}$ contain
the effect of the decay process $\Delta\rightarrow\pi N$ or $\rho
\rightarrow\pi\pi$ and their explicit forms are given in Ref.\cite{ochi1}.

In this paper, we simply choose 400 MeV/c for the range parameters ($p_{\pi
NN^{\ast}}$, $p_{\pi\Delta N^{\ast}}^{s}$, $p_{\pi\Delta N^{\ast}}^{d}$,
$p_{\rho NN^{\ast}}$) which reproduces the nucleon size in quark
models\cite{arima}. The coupling constants ($f_{\pi NN^{\ast}}$, $f_{\pi\Delta
N^{\ast}}^{s}$, $f_{\pi\Delta N^{\ast}}^{d}$, $f_{\rho NN^{\ast}}$) and the
bare mass ($M_{N^{\ast}}^{0}$) are adjusted to fit the $N^{\ast}$ resonance
energy, its width and the branching ratios at the resonance energy. We use
1520 MeV as the resonance energy and 120 MeV as the width, respectively. We
take a fraction of 60\% for the decay into $\pi N$, 8\% into s-wave $\pi
\Delta$, 12\% into d-wave $\pi\Delta$ and 20\% decay into the $\rho N$
channel, respectively\cite{PDG}, which are slightly different from the values
used in the previous model\cite{ochi1,ochi2}. We note that the parameters 
cannot be uniquely fixed due to the limited experimental information and their
uncertainties. The parameter set used in this paper are given in Table 1.
\ The signs of the coupling constants are the same as those in
Ref.\cite{ochi2}. With this parameterization, the $\pi N$ D$_{13}$scattering
amplitudes calculated agree with the data around the resonance energy but are
deviated as the pion energy is away from it. The range parameters are
necessary to be varied in order to get good agreement with the data over a
wide range of the pion energies, but they are not uniquely determined in the
present case and the fitted parameter sets include a very small range
parameter which is hardly acceptable from a physical point of view.
Furthermore, the background $\pi N$ interaction may be needed and thus one
must go beyond the present framework of the isobar model. We think that the
above description for $N^{\ast}$ is sufficient for the present purpose,
namely, to examine the difference between the PS and PV couplings.%

\begin{table}[hbtp] \centering
\caption{Parameter set for $N^{*}$ used in this paper\label{Table 1}}
\begin{tabular}
[c]{|l|l|l|l|l|}\hline
$f_{\pi NN^{\ast}}$ & $f_{\pi\Delta N^{\ast}}^{s}$ & $f_{\pi\Delta N^{\ast}%
}^{d}$ & $f_{\rho NN^{\ast}}$ & $M_{N^{\ast}}^{0}(MeV)$\\\hline
1.147 & 0.398 & 1.435 & 0.942 & \ \ 1709.\\\hline
\end{tabular}
\end{table}%

Now we consider the resonant couplings by photon. The $\gamma N\Delta$ and
$\gamma NN^{\ast}$ coupling constants can be determined by using the multipole
amplitudes in the relevant channel for the $\gamma N\rightarrow\pi N$
reaction. The multipole amplitudes for the $\gamma N\rightarrow\pi N$ reaction
have non-negligible background contributions and thus are generally expressed
as the sum of the background and resonant terms, i.e.,%

\begin{equation}
T^{\gamma N}=T_{B}^{\gamma N}+T_{R}^{\gamma N}.
\end{equation}
Like the $\pi N$ elastic scattering amplitudes, the resonant term
$T_{R}^{\gamma N}$ is given by the isobar model as%

\begin{equation}
T_{\Delta}^{\gamma N}=\frac{F_{\pi N\Delta}F_{\gamma N\Delta}^{\dag}}{\sqrt
{s}-M_{\Delta}^{0}-\Sigma_{\pi N}^{\Delta}},
\end{equation}
for the $\Delta$ resonance and%

\begin{equation}
T_{N^{\ast}}^{\gamma N}=\frac{F_{\pi NN^{\ast}}F_{\gamma NN^{\ast}}^{\dag}%
}{\sqrt{s}-M_{N^{\ast}}^{0}-\Sigma_{\mathrm{total}}^{N^{\ast}}},
\end{equation}
for the $N^{\ast}$ resonance.

The $\Delta$ resonance can contribute to both $M_{1+}(3/2)$ and $E_{1+}(3/2)$
multipole amplitudes. Since the magnitude of the $E_{1+}(3/2)$ multipole is
small compared with $M_{1+}(3/2)$, the $E2$ $\gamma N\Delta$ coupling is
neglected. The $\gamma N\rightarrow\Delta$ vertex function for the
$M_{1+}(3/2)$ channel is written as%

\begin{equation}
F_{\gamma N\Delta}^{\dag}=-ig_{M_{1+}}\mathbf{S}^{\dagger}\cdot\mathbf{k}%
\times\bm{\epsilon}  ,
\end{equation}
where $g_{M_{1+}}$ and $\bm{\epsilon}  $ are the $M1$ $\gamma N\Delta$
coupling constant and photon polarization vector, respectively, and
$\mathbf{k}$ denotes the initial photon momentum. We use $g_{M_{1+}}=0.1991$
(in natural unit), which is obtained from the resonant coupling given by the
Particle Data Group\cite{PDG}.

For the $N^{\ast}$ resonance, we use the helicity amplitudes instead of the
electric and magnetic multipoles. The $\gamma NN^{\ast}$ vertex has two
independent helicity couplings: the helicity 1/2 and 3/2 couplings. For the
helicity 1/2 transition, $F_{\gamma NN^{\ast}}^{\dag}$ is written as
\begin{equation}
F_{\gamma NN^{\ast}}^{1/2\dag}=-ig_{1/2}\left(  \mathbf{S}^{\dagger}%
\cdot\widehat{\mathbf{k}}\right)  (\bm{\sigma}{\cdot}  \widehat{\mathbf{k}%
}\times\bm{\epsilon}  ),
\end{equation}
where $g_{1/2}$ is the helicity 1/2 coupling constant and $\widehat
{\mathbf{k}}$ denotes the unit vector of the initial photon momentum. For the
helicity 3/2 transition, $F_{\gamma NN^{\ast}}^{\dag}$ is written as
\begin{equation}
F_{\gamma NN^{\ast}}^{3/2\dag}=g_{3/2}\left\{  \left(  \mathbf{S}^{\dagger
}\cdot\bm{\epsilon}  \right)  +\frac{i}{2}\left(  \mathbf{S}^{\dagger}%
\cdot\widehat{\mathbf{k}}\right)  (\bm{\sigma}{\cdot}  \widehat{\mathbf{k}%
}\times\bm{\epsilon}  )\right\}  ,
\end{equation}
where $g_{3/2}$ is the helicity 3/2 coupling constant. For the proton target,
the helicity 1/2 amplitude is small compared with the helicity 3/2 amplitude
and so the helicity 1/2 coupling is neglected. For the neutron target, both
couplings are taken into account. We use $g_{1/2}=0$ and $g_{3/2}=0.1612$ for
the proton target and $g_{1/2}=-0.0496$ and $g_{3/2}=-0.135$ for the neutron
target, respectively, which are obtained from the resonant couplings given by
the Particle Data Group\cite{PDG}.

Generally, the $\gamma N\Delta$ and $\gamma NN^{\ast}$ coupling constants (
$g_{M_{1+}}$, $g_{1/2}$ , $g_{3/2}$) include both bare couplings to the
resonances and vertex corrections due to the interference 
with the background processes and thus are complex and energy dependent in nature.
One
way to determine the coupling constants is to extract them from the
experimental multipole amplitudes by assuming an appropriate non-resonant
background term $T_{B}^{\gamma N}$. The other way is to use the resonance
couplings given by the Particle Data Group \cite{PDG} which correspond to the
bare couplings. In most of previous models\cite{laget,oset1,nacher1}, real
coupling constants obtained by the latter method have been used to calculate
the cross sections for the two-pion photoproduction. Since the imaginary
values are small for these resonances as predicted in the phenomenological
calculations\cite{tiator}, we also employed the latter method for simplicity.

\subsection{resonant amplitude of $\gamma N\rightarrow\pi\pi N$}

We use the same approach with the model of Ref.\cite{ochi1,ochi2} for the
resonance production processes and make a brief review about it in this
subsection. The resonant $T_{R}$ matrix for the two-pion photoproduction is
expressed as
\begin{equation}
T_{R}=T_{\Delta KR}+T_{\Delta PP}+T_{N^{\ast}\pi\Delta}^{s}+T_{N^{\ast}%
\pi\Delta}^{d}+T_{N^{\ast}\rho N}+T_{\Delta nex}. \label{tr}%
\end{equation}
The resonant $T_{R}$ matrix consists of six amplitudes: the $\Delta$
Kroll-Ruderman term ($T_{\Delta KR}$), $\Delta$ pion-pole term ($T_{\Delta
PP}$), $N^{\ast}$ excitation terms ($T_{N^{\ast}\pi\Delta}^{s(d)}$ and
$T_{N^{\ast}\rho N}$) and the $\pi\Delta$ production term accompanied by
nucleon exchange ($T_{\Delta nex}$). These diagrams are shown in Fig. 5 (a)-(e),
respectively. The strength of these resonant processes depends strongly on the
isospin channel\cite{oset1,ochi1}. The terms $T_{\Delta KR}$ and $T_{\Delta
PP}$ dominate for the $\pi^{+}\pi^{-}$ production and becomes small due to the
isospin factor for the $\pi^{+}\pi^{0}$($\pi^{-}\pi^{0}$) production and then
are prohibited for the $\pi^{0}\pi^{0}$ production. Although the $N^{\ast}$
excitation terms have only weak strength, they contribute to all isospin
channel and their interference with the $\Delta$ Kroll-Ruderman term has
significant effects to the cross section. The other processes arising from the
requirement of the gauge invariance are neglected because they are known to be
small.%
\begin{figure}
[ptb]
\begin{center}
\includegraphics[
trim=0.673883in 4.412486in 0.681325in 0.894660in,
height=4.3543in,
width=4.184in
]%
{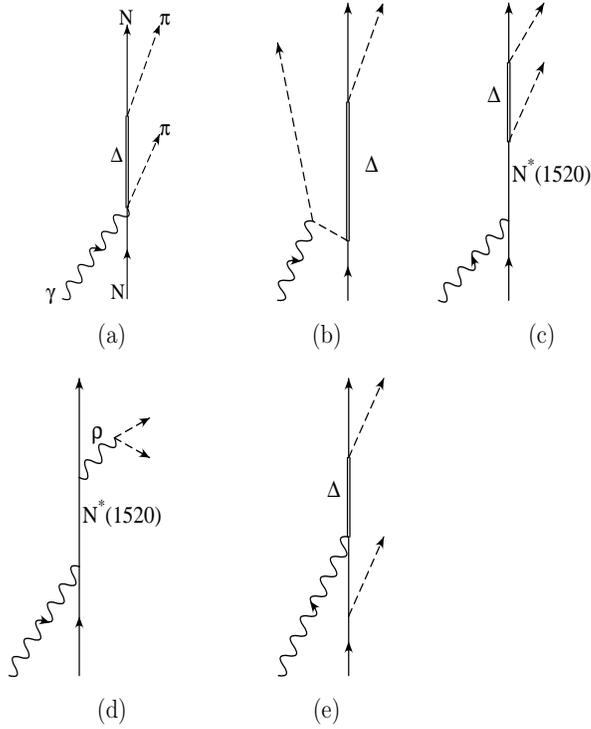}%
\caption{Diagrams of resonant processes.(a) the $\Delta$ Kroll-Ruderman term
($T_{\Delta KR}$), (b) $\Delta$ pion-pole term ($T_{\Delta PP}$), (c)
$N^{\ast}$ excitation terms ($T_{N^{\ast}\pi\Delta}^{s(d)}$), (d) $N^{\ast}$
excitation terms ($T_{N^{\ast}\rho N}$) and (e) the $\pi\Delta$ production
term accompanied by nucleon exchange ($T_{\Delta nex}$).}%
\label{fig5}%
\end{center}
\end{figure}

The $\Delta$ Kroll-Ruderman term is written as%

\begin{equation}
T_{{\Delta\mathrm{KR}}}=\frac{F_{\pi N\Delta}F_{\Delta\mathrm{KR}}^{\dag}%
}{\sqrt{s}-\omega_{\pi}-E_{\Delta}-\Sigma_{\Delta}^{(\pi N)}(q,\sqrt{s})},
\label{tkr}%
\end{equation}
where $\omega_{\pi}=\sqrt{m_{\pi}^{2}+q^{2}}$ and $E_{\Delta}=\sqrt
{(M_{\Delta}^{0})^{2}+q^{2}}$. $\Sigma_{\Delta}^{(\pi N)}(q,\sqrt{s})$ is the
self-energy of $\Delta$ moving with the momentum $q$ that arises from the
coupling to the $\pi N$ channel. Its expression is given in Ref. \cite{arima}.
The $\gamma N\pi\Delta$ contact term operator $F_{\Delta\mathrm{KR}}^{\dag}$
is obtained from the strong $\pi N\Delta$ vertex function so as to satisfy the
gauge invariance, the detail of which is given in Ref.\cite{ochi1}. Instead of
using the effective Lagrangian\cite{oset1}, we employ the vertex function with
a form factor. The $N\rightarrow\pi\Delta$ transition operator is assumed to
have the same form as the $\Delta\rightarrow\pi N$ vertex function. 
Since the range parameter $Q_{\Delta
}(N\rightarrow\pi\Delta)$ is not necessarily the same as the parameter
$Q_{\Delta}$ in Eq.(\ref{qdelta}) and thus is unknown, we treat it as a free
parameter and vary it to fit the $\gamma p\rightarrow\pi^{+}\pi^{-}p$ cross
section. The pion-pole term $T_{\Delta PP}$ is obtained by replacing
$F_{\Delta\mathrm{KR}}^{\dag}$ in Eq.(\ref{tkr}) with the $\gamma N\pi\Delta$
pion-pole vertex function $F_{\Delta\mathrm{PP}}^{\dag}$ whose detailed
expression is given in Ref\cite{ochi1}.

The $\gamma N\rightarrow N^{\ast}\rightarrow\pi\pi N$ transition takes place
the following two possible processes: $N^{\ast}\rightarrow\pi\Delta$ (s-wave
or d-wave $\pi\Delta$ state) and $N^{\ast}\rightarrow\rho N$. They are
described by $T_{N^{\ast}\pi\Delta}^{s(d)}$ and $T_{N^{\ast}\rho N}$,
respectively. Using the isobar model mentioned in the previous subsection, the
T matrix elements of $T_{N^{\ast}\pi\Delta}^{s(d)}$ and $T_{N^{\ast}\rho N}$
are written as%

\begin{align}
T_{N^{\ast}\pi\Delta}^{\mathrm{s(d)}}  &  =\frac{F_{\pi N\Delta}F_{\pi\Delta
N^{\ast}}^{s(d)\dag}F_{\gamma NN^{\ast}}^{\dag}}{(\sqrt{s}-\omega_{\pi
}-E_{\Delta}-\Sigma_{\Delta}^{(\pi N)}(q,\sqrt{s}))\left(  \sqrt{s}%
-M_{N^{\ast}}^{0}-\Sigma_{\mathrm{total}}^{N^{\ast}}\right)  },\\
T_{N^{\ast}\rho N}  &  =\frac{F_{\rho\pi\pi}F_{\rho NN^{\ast}}F_{\gamma
NN^{\ast}}^{\dag}}{2\omega_{\rho}\left(  \sqrt{s}-\omega_{\rho}-E_{q_{\rho}%
}-\Sigma_{\rho\pi\pi}(q_{\rho},\sqrt{s})\right)  \left(  \sqrt{s}-M_{N^{\ast}%
}^{0}-\Sigma_{\mathrm{total}}^{N^{\ast}}\right)  },
\end{align}
respectively. Here, $q_{\rho}$ is the momentum of $\rho$ meson and
$\Sigma_{\rho\pi\pi}(q_{\rho},\sqrt{s})$ is the self-energy of $\rho$ meson
moving with the momentum $q_{\rho}$. The $\rho$ meson is described by the
isobar model (see Eq.(\ref{rho-isobar})).

The $\pi\Delta$ production term accompanied by nucleon exchange is written as%

\begin{equation}
T_{\Delta nex}=\frac{F_{_{\pi N\Delta}}F_{\gamma N\Delta}^{\dagger}F_{\pi
NN}^{\dagger}}{(\sqrt{s}-\omega_{\pi}-E_{\Delta}-\Sigma_{\Delta}^{(\pi
N)}(q,\sqrt{s}))(E_{k}-\omega_{\pi}-E_{k+q})}, \label{dnex}%
\end{equation}
where the $\pi NN$ vertex function $F_{\pi NN}^{\dagger}$ is assumed to have a
usual non-relativistic form. Unlike the other resonant processes, the
intermediate particle, i.e., nucleon, is far off-shell. Since only the
on-shell resonant $M_{1+}(3/2)$ multipole amplitude is known, some
prescription is needed to include the off-shell effect in the above
T matrix (Eq.(\ref{dnex})). We take the modified pole approximation
\cite{ferrari}where the angular part of $F_{\gamma N\Delta}^{\dagger}$ is
evaluated at the center of mass system of $\gamma N$ and its magnitude is
evaluated at the total energy of the final $\pi N$ state. The contribution of
$T_{\Delta nex}$ to the cross sections is expected to be small due to the
off-shell effect. However we include this amplitude in our calculations, since
it becomes non-negligible for the $\gamma p\rightarrow\pi^{0}\pi^{0}p$
reaction due to large coupling constants in $M_{1+}(3/2)$ channel.

\section{Numerical results and discussions}

In this section, we present our numerical results of total cross sections,
invariant mass spectra and helicity-dependent cross sections obtained by using
the model introduced in the previous sections and compare them with the
experimental data. In our model, the T matrix for the two-pion photoproduction
is written as
\begin{equation}
T=T_{NR}+T_{R}, \label{T}%
\end{equation}
where $T_{NR}$ is the non-resonant T matrix given in section II and $T_{R}$ is
the resonant T matrix (Eq.(\ref{tr})) in section III. For $T_{NR}$, we
consider two kinds of models: PS and PV models with vector meson contributions.
Correspondingly there
are two kinds of full T matrices for Eq.(\ref{T}) which are expressed as
$T_{PS}$ and $T_{PV}$, respectively. To demonstrate the difference of $\pi NN$
couplings, we always compare the calculations by $T_{PS}$ with those by
$T_{PV}$ which are shown in figures below. \ 

In our model, there is a free parameter: the range parameter $Q_{\Delta
}(N\rightarrow\pi\Delta)$ of the $N\rightarrow\pi\Delta$ vertex function
appearing in the $\Delta$ Kroll-Ruderman term. At present we do not know how
to determine it by using other reactions than the two-pion photoproduction. We
use $Q_{\Delta}(N\rightarrow\pi\Delta)$ =430 MeV/c determined so as to
reproduce the $\gamma p\rightarrow\pi^{+}\pi^{-}p$ cross sections. Here the
$N\rightarrow\pi\Delta$ vertex function is assumed to have the same form with
the $\Delta\rightarrow\pi N$ vertex function used in this paper, since the
$\Delta$ propagator used in Eq.(\ref{tkr}) is calculated with the same
$\Delta\rightarrow\pi N$ vertex function. In this sense, we think, it is not
appropriate to use the monopole type of form factor employed by the authors of
Refs.\cite{oset1,nacher1} in our approach. Although the value of $Q_{\Delta
}(\Delta\rightarrow\pi N)$ is taken to be 358 MeV/c given by Betz and
Lee\cite{betz}, \ the value of $Q_{\Delta}(N\rightarrow\pi\Delta)$ does not
need to be the same, since the form factors for $\Delta\rightarrow\pi N$ and
$N\rightarrow\pi\Delta$ are functions depending on the relative momentum of
$\pi N$ and $\pi\Delta$ systems, respectively. At a given relative momentum,
in fact, the square of the pion four-monentum evaluated in the $\pi N$ center
of mass system is larger than that in the $\pi\Delta$ center of mass system
due to the mass difference between $N$ and $\Delta.$

\begin{figure}[ptb]
\begin{center}
\scalebox{0.8}{\includegraphics{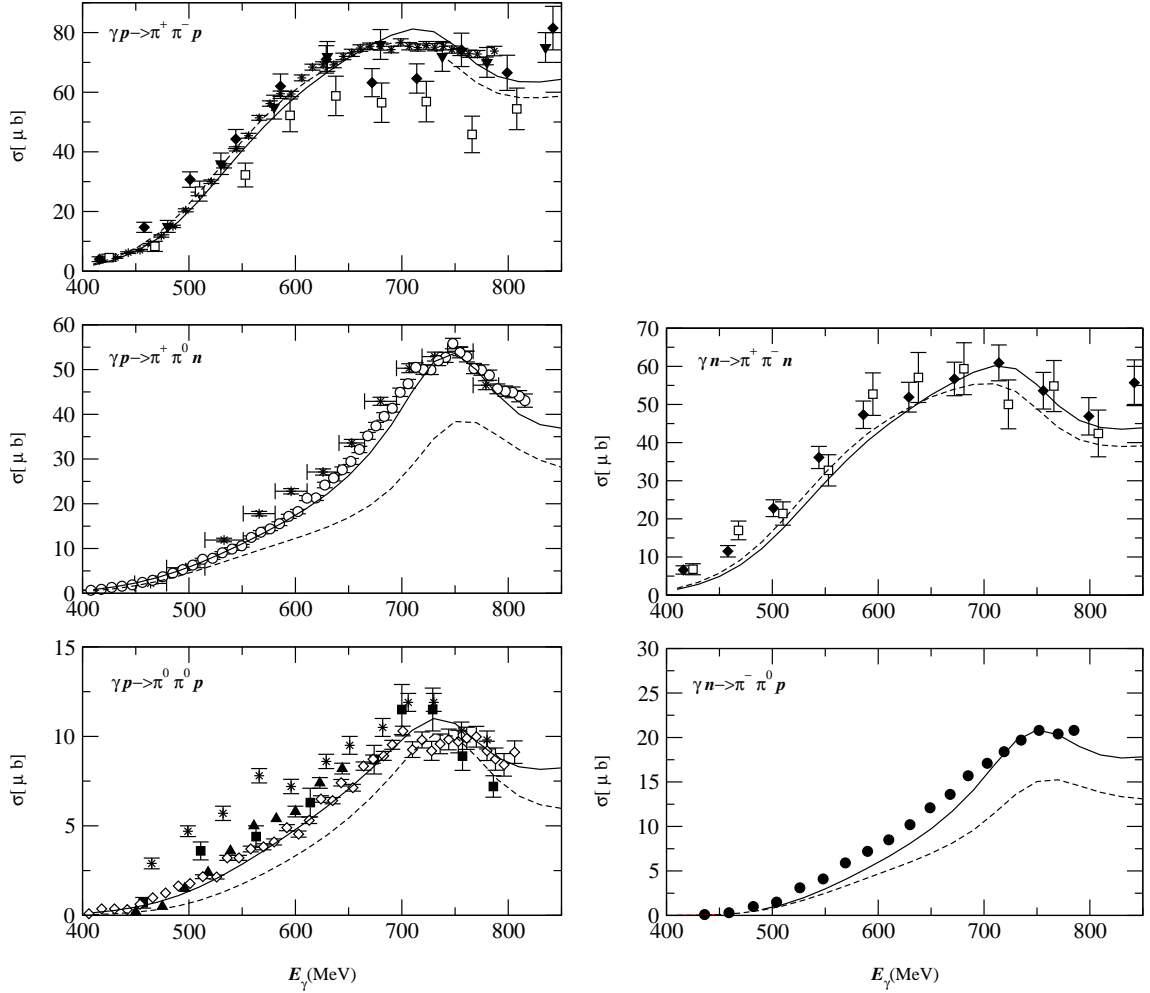}}
\end{center}
\caption{{\small {Total cross sections of two-pion photoproduction on proton
and neutron in various isospin channels.The solid and dashed lines correspond
to the PS and PV calculations. Data are from Refs.\cite{ABB}(triangle down),
\cite{pia}(open square),\cite{car}(black diamond),\cite{bra}(star),
\cite{lan}(open circle),\cite{hae}(triangle up and black square),
\cite{wolf}(open diamond) and \cite{zab}(black circle) . The data (black
circle) of $\gamma n\rightarrow\pi^{-}\pi^{0}p$ correspond to the cross
section over the DAPHNE acceptance. }}}%
\label{fig6}%
\end{figure}

First, we show the results of the total cross sections of five isospin
channels, i.e., $\gamma p\rightarrow\pi^{+}\pi^{-}p$, $\gamma p\rightarrow
\pi^{+}\pi^{0}n$, $\gamma p\rightarrow\pi^{0}\pi^{0}p$, $\gamma n\rightarrow
\pi^{+}\pi^{-}n$ and $\gamma n\rightarrow\pi^{-}\pi^{0}p,$\ in Fig.6. Here the
calculations with $T_{PS}$ and $T_{PV}$ are presented with the solid and
dashed lines, respectively. We find that the PS calculations are in good
agreement with the data in all channels and on the other hand the PV
calculations underestimate the data except for $\pi^{+}\pi^{-}$ channels. The
discrepancy between two calculations is clearly due to the difference of the
non-resonant processes. Our results of the $\gamma p\rightarrow\pi^{+}\pi
^{0}n$ and $\gamma p\rightarrow\pi^{0}\pi^{0}p$ reactions for the PV model are
not consistent with those by the model of Nacher et al.\cite{nacher1} where
the non-resonant amplitudes are constructed using the non-relativistic PV $\pi
NN$ coupling. Their results are more close to the experimental data compared
with our result for the PV model. We think that this inconsistency comes
mainly from the way of the calculation of the diagram (e) in Fig.5, where the
nucleon in the intermediate state is far off-shell. We have evaluated this
diagram with the modified pole approximation and on the other hand they
adopted simply the $\gamma N\Delta$ vertex function without cut-off factor
which has a linear dependence of the photon momentum. Consequently, the
$\gamma N\Delta$ coupling becomes very large in the energy region of the
$N^{\ast}(1520)$ resonance and leads to the large cross section.%
\begin{figure}
[ptb]
\begin{center}
\includegraphics[
height=2.9568in,
width=6.2301in
]%
{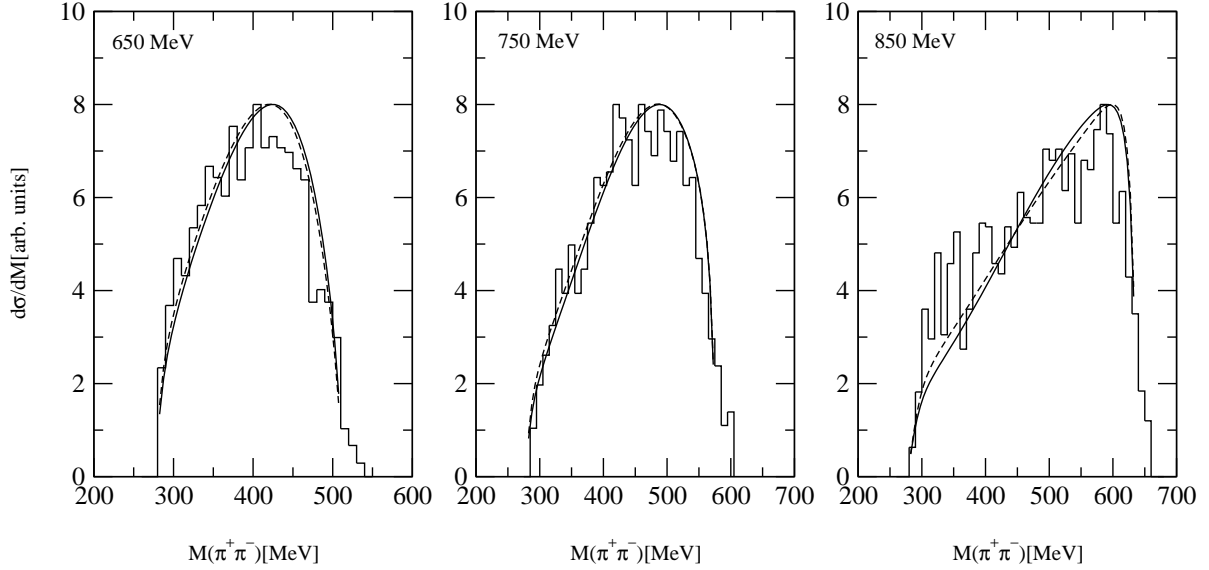}%
\caption{Invariant mass spectra of $\pi\pi$\ for $\gamma p\rightarrow\pi
^{+}\pi^{-}p$ at 650, 750 and 850 MeV. The solid and dashed lines correspond
to the PS and PV calculations, respectively. Data are normalized appropriately
and from Ref.\cite{ABB}.}%
\label{fig7}%
\end{center}
\end{figure}

Secondly, we calculated the invariant mass spectra of three isospin channels,
i.e., $\gamma p\rightarrow\pi^{+}\pi^{-}p$, $\gamma p\rightarrow\pi^{+}\pi
^{0}n$ and $\gamma p\rightarrow\pi^{0}\pi^{0}p$. The invariant mass spectra
for the $\pi^{+}\pi^{-}$ system in the $\gamma p\rightarrow\pi^{+}\pi^{-}p$
channel are shown in Fig.7. The calculations are performed with three photon
energies, i.e., 650, 750 and 850 MeV. In this case, the values of the data are
plotted arbitrarily and the calculations are normalized so as to fit the peak
value of the experimental distributions. For the shape, two calculations (PS
and PV calculations) are almost equivalent each other and are in good
agreement with the data.

\begin{figure}[ptb]
\begin{center}
\scalebox{0.7}{\includegraphics{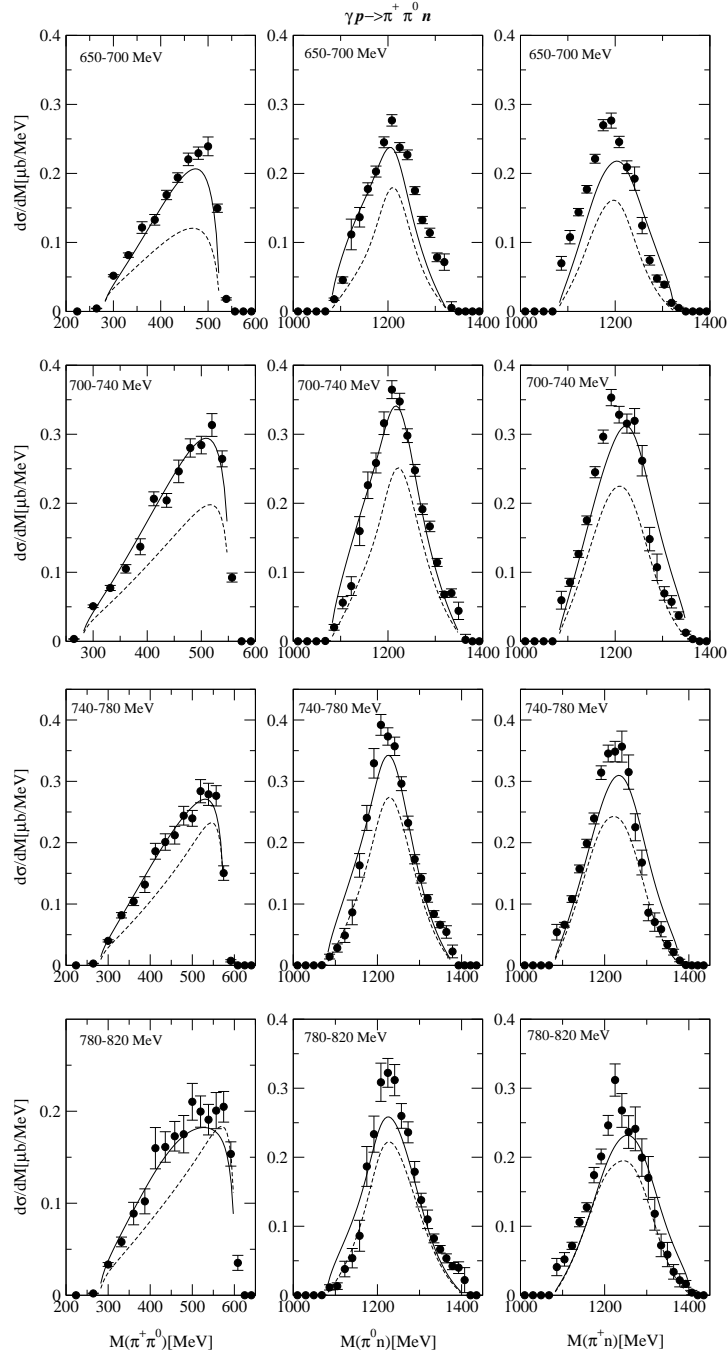}}
\end{center}
\caption{Invariant mass spectra of $\pi^{+}\pi^{0}$, $\pi^{0}n$ and $\pi^{+}n$
for $\gamma p\rightarrow\pi^{+}\pi^{0}n$ at four bins of incident photon
energy. The solid and dashed lines correspond to the PS and PV calculations,
respectively. Data are from Ref.\cite{lan}}%
\label{fig8}%
\end{figure}

For the $\gamma p\rightarrow\pi^{+}\pi^{0}n$ channel, the invariant mass
spectra for the $\pi^{+}\pi^{0}$, $\pi^{0}n$ and $\pi^{+}n$ systems are
calculated at four bins of photon energies, i.e., 650-700 MeV, 700-740 MeV,
740-780MeV and 780-820 MeV and are shown with the data in Fig.8. \ We find
that the PS calculations agree well with all data but the PV calculations show
some discrepancy about the shape of the $\pi^{+}\pi^{0}$ invariant mass
spectra in addition to the magnitude. The $\pi^{+}\pi^{0}$ invariant mass
spectra have a peak shifted to the higher $\pi\pi$ invariant mass and the peak
position of the $\pi^{0}n$ and $\pi^{+}n$ invariant mass spectra corresponds
to the mass of $\Delta(1232)$. In order to see which process makes such
behavior, the contributions of some components of the resonant processes are
shown in Fig.9.
\begin{figure}
[ptb]
\begin{center}
\includegraphics[
height=3.1453in,
width=6.2301in
]%
{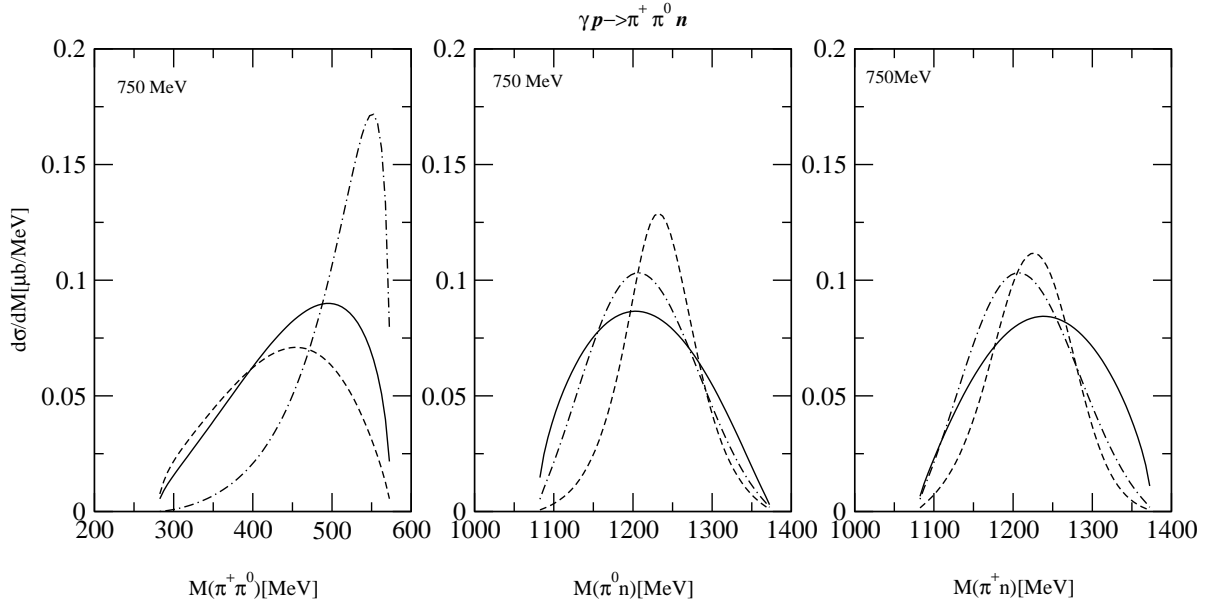}%
\caption{Contributions of resonant processes in invariant mass spectra of
$\pi^{+}\pi^{0}$, $\pi^{0}n$ and $\pi^{+}n$ systems for $\gamma p\rightarrow
\pi^{+}\pi^{0}n$ at 750 MeV. The dash-dotted and dashed lines are the
calculations with $T_{N^{\ast}\rho N}$ and $T_{\Delta KR}+T_{\Delta
PP}+T_{\Delta nex}+T_{N^{\ast}\pi\Delta}^{s}+T_{N^{\ast}\pi\Delta}^{d}$(see
text), respectively. The non-resonant PS calculation (solid line) is also
plotted.}%
\label{fig9}%
\end{center}
\end{figure}
Here the dash-dotted and dashed lines correspond to the calculations with
$T_{N^{\ast}\rho N}$ and $T_{\Delta KR}+T_{\Delta PP}+T_{\Delta nex}%
+T_{N^{\ast}\pi\Delta}^{s}+T_{N^{\ast}\pi\Delta}^{d}$, respectively. For
reference, the non-resonant PS calculations (solid lines) are also plotted in
figures. Clearly, the peak shift for the $\pi^{+}\pi^{0}$ invariant mass
distribution is due to the contribution of $N^{\ast}\rightarrow\rho N$ process
and the peak position of the $\pi^{0}n$ and $\pi^{+}n$ invariant mass
distribution is directly related to the $\Delta(1232)$ production in the
intermediate state.

\begin{figure}[ptb]
\begin{center}
\scalebox{0.7}{\includegraphics{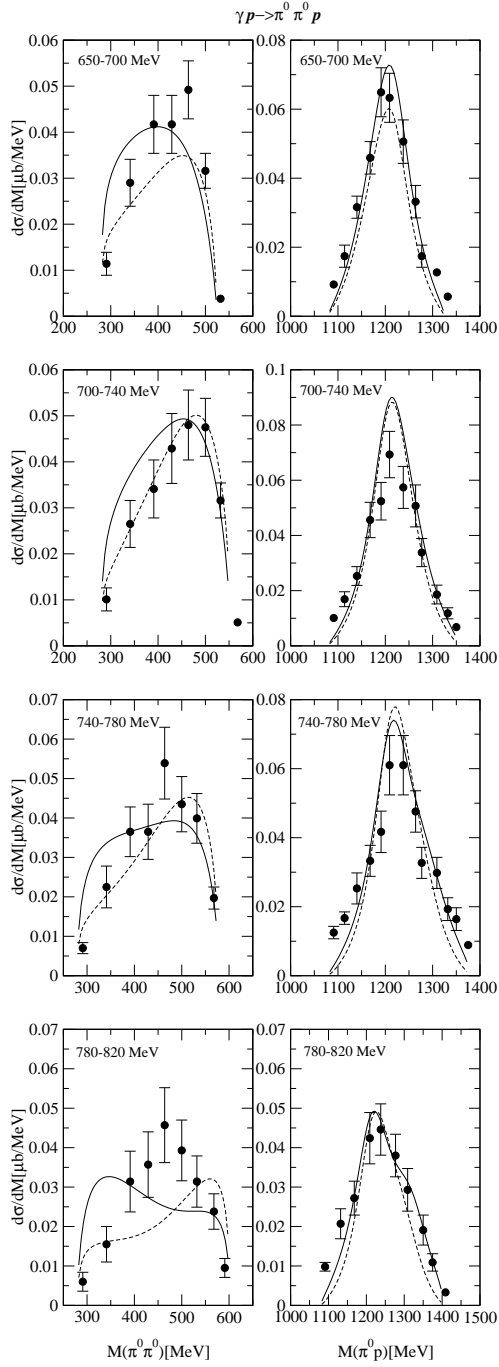}}
\end{center}
\caption{Invariant mass spectra of $\pi^{0}\pi^{0}$ and $\pi^{0}p$ for $\gamma
p\rightarrow\pi^{0}\pi^{0}p$ at four bins of incident photon energy. The solid
and dashed lines correspond to the PS and PV calculations, respectively. Data
are from Ref.\cite{wolf}}%
\label{fig10}%
\end{figure}

For the $\gamma p\rightarrow\pi^{0}\pi^{0}p$ channel, the invariant mass
spectra for the $\pi^{0}\pi^{0}$ and $\pi^{0}p$ systems are calculated at the
same bins of photon energies with the $\gamma p\rightarrow\pi^{+}\pi^{0}n$
channel. The results are shown with the data in Fig.10. In this channel, the
$\Delta$ Kroll-Ruderman term (Fig.5(a)) and pion pole terms (Fig.5(b)) do not
contribute to the cross section and so the magnitude of the cross section is
rather small compared with other channels. Furthermore the $I=J=1$ $\pi\pi$
system such as the $\rho$ meson \ is not produced because of isospin
conservation. Therefore only the processes of $T_{\Delta nex}$,$T_{N^{\ast}%
\pi\Delta}^{s}$ and $T_{N^{\ast}\pi\Delta}^{d}$ among the resonant processes
contribute to the cross section in our model. We note that the production of
the $I=J=0$ $\pi\pi$ system such as the $\sigma$ meson could take place in
this channel. From the comparison with the data, one finds that both PS and PV
calculations can almost reproduce the data of the $\pi^{0}p$ invariant mass
spectra, which has a peak at the $\Delta(1232)$ mass. For the $\pi^{0}\pi^{0}$
invariant mass spectra, there are some discrepancies, especially, between PS
calculations and the data. The PV calculations are almost consistent with the
data except for the bin of 780-820 MeV, where there are two bumps in the
distribution. On the other hand, the PS calculations show broader
distributions than the data, arising from the non-resonant processes as shown
in Fig.4. In our calculations, the final state interactions for $\pi^{0}%
\pi^{0}$ and $\pi^{0}p$ are not taken into account. As pointed out in
Ref.\cite{oset-pipi}, the final state interaction for $\pi^{0}\pi^{0}$ is
expected to be important because of strong correlation between two pions in
$I=J=0$ channel and the influence of the $\sigma$ meson. Such effect might be
one of the possibilities to improve the PS model and\ so it will be worthwhile
to further study this channel by taking into account such effect.

\begin{figure}[ptb]
\begin{center}
\scalebox{0.5}{\includegraphics{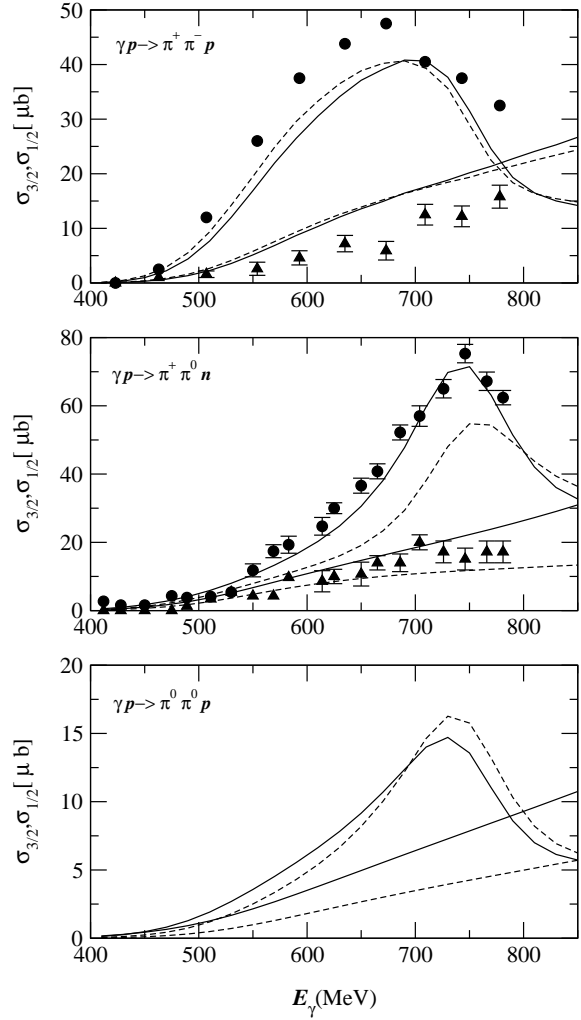}}
\end{center}
\caption{Helicity-dependent cross sections $\sigma_{3/2}$ and $\sigma_{1/2}$
for $\gamma p\rightarrow\pi^{+}\pi^{-}p$, $\gamma p\rightarrow\pi^{+}\pi^{0}n$
and $\gamma p\rightarrow\pi^{0}\pi^{0}p$. The solid and dashed lines
correspond to PS and PV calculations, respectively. The upper two lines and
black circles correspond to the cross section $\sigma_{3/2}$ and the other two
lines and triangles correspond to the cross section $\sigma_{1/2}$ in each
figure. In the above two figures of $\gamma p\rightarrow\pi^{+}\pi^{-}p$ and
$\gamma p\rightarrow\pi^{+}\pi^{0}n$ , the cross sections over the DAPHNE
acceptance are shown. Data are from Ref.\cite{nacher2,lang}}%
\label{fig11}%
\end{figure}

Finally the calculations of the helicity-dependent cross sections
$\sigma_{3/2}$ and $\sigma_{1/2}$ are shown with the data in Fig. 11. Here we
display the cross sections in three isospin channels: the $\gamma
p\rightarrow\pi^{+}\pi^{-}p$\ cross section with the three charged particles
in the DAPHNE acceptances\cite{nacher2,lang}, the $\gamma p\rightarrow\pi
^{+}\pi^{0}n$ cross section with the $\pi^{+}$ in the DAPHNE
acceptance\cite{nacher2,lang} and the $\gamma p\rightarrow\pi^{0}\pi^{0}p$
cross section without kinematical limits for outgoing particles. The quantity
$\sigma_{3/2}$ ($\sigma_{1/2}$) is defined as the cross section for the
absorption of a polarized photon by a polarized target proton in the helicity
3/2 (1/2) channel. The solid and dashed lines denote the calculations with the
PS model and PV model, respectively. It should be noted that the
electromagnetic coupling of the $N^{\ast}(1520)$ to proton is known to be
dominated by the helicity 3/2 state\cite{PDG}. This is why the peak of the
resonance is seen in the data for $\sigma_{3/2}$ but not for $\sigma_{1/2}$.
Therefore the cross section $\sigma_{1/2}$ is sensitive to the reaction
mechanism not related to the $N^{\ast}(1520)$ resonance. For the $\gamma
p\rightarrow\pi^{+}\pi^{-}p$ reaction, one finds that the PS calculation is
almost equivalent with the PV calculation like the total cross sections.
Although the calculations agree qualitatively with the data, the cross
section $\sigma_{3/2}$ is underestimated and the cross section $\sigma_{1/2}$
is overestimated. The detailed description of the reaction mechanism is still
unsatisfactory for this channel and is needed to pursue what is missing in our
model. For the $\gamma p\rightarrow\pi^{+}\pi^{0}n$ reaction, the PS model can
almost explain two helicity-dependent cross sections simultaneously except for
the cross section $\sigma_{1/2}$ around 750 MeV. On the other hand, the PV
model underestimates the cross section $\sigma_{3/2}$ and also $\sigma_{1/2}$
slightly. In this model, non-resonant processes in the helicity 3/2 channel
are not strong enough to explain the data. This result is not consistent with
that obtained from the model by Nacher et al.\cite{nacher2}. This is due to
the same reason mentioned in the discussions of the total cross sections. For
the $\gamma p\rightarrow\pi^{0}\pi^{0}p$ reaction, we observe that there is a
large difference between the PS and PV models in the cross section
$\sigma_{1/2}$. The experimental data, if exist, could provide an important
information on the reaction mechanism.

\section{Concluding remarks}

Several theoretical studies on the two-pion photoproduction have been
performed in the past but none of them have succeeded to reproduce
the data in all isospin channels simultaneously. Particularly, unexpectedly
large cross sections of the $\gamma p\rightarrow\pi^{+}\pi^{0}n$ and $\gamma
n\rightarrow\pi^{-}\pi^{0}p$ reactions were found not to be explained with the
usual reaction mechanism and thus the presence of a new reaction
mechanism in these channels is suggested. Ochi et al.\cite{ochi1}introduced the $\rho$
Kroll-Ruderman term, which influences only on the above isospin channels, as a
new reaction mechanism. However, a soft $\rho\pi\pi$ form factor was needed to
reproduce the large cross sections and the isobar model with such a form
factor failed to explain the $\pi\pi$ scattering at low energies. Thus, the
$\rho$ Kroll-Ruderman term was inferred to represent the effect of the
background process followed by the production of the $I=J=1$ $\pi\pi$ system
rather than the $\rho$ meson itself.

In this paper we have discussed the effect of the non-resonant processes
arising from the PS and PV $\pi NN$ couplings in the two-pion photoproduction
in order to pursue an alternative reaction mechanism. The non-resonant
amplitudes can be obtained by attaching an external photon line to the diagram
of $N\rightarrow\pi\pi N$ calculated with the two $\pi NN$ couplings. In order
that the model with the PS $\pi NN$ coupling is equivalent to the model with 
the PV $\pi NN$ coupling 
in the strong interaction process $N\rightarrow\pi\pi N$, \ the
effective contact interaction (Eq.(\ref{contact})) is added to the former model. 
Using these models, we examined the effect of the
two couplings on various isospin channels numerically. For the $\gamma
p\rightarrow\pi^{+}\pi^{0}n$ and $\gamma n\rightarrow\pi^{-}\pi^{0}p$
reactions, we found that the cross sections calculated with the PS model were
larger than those with the PV model and their magnitude was almost consistent
with the $\rho$ Kroll-Ruderman\ used in Ref.\cite{ochi1}. Consequently,
the non-resonant process described by the PS $\pi NN$ coupling can be regarded
as a candidate for the new reaction mechanism in place of the $\rho$
Kroll-Ruderman term.

The importance of the PS $\pi NN$ coupling in the two-pion photoproduction has
not been so far noticed. In fact the non-resonant process has been always
described by the PV $\pi NN$ coupling and has been found to have only a minor
contribution to the two-pion photoproduction. The PV $\pi NN$ coupling is more
favored than the PS $\pi NN$ coupling for the $\gamma N\rightarrow\pi N$
reaction and the $\pi N$ scattering at low energies. Therefore, the PV coupling
has been also used for the $\gamma N\rightarrow\pi\pi N$ reaction as a matter
of course. Recently, Drechsel et al. \cite{tiator} pointed out in the study of
the $\gamma N\rightarrow\pi N$ reaction that the PS coupling was needed to
describe the data with the increase of the incident photon energy. This
implies that the PS coupling is preferable to the PV coupling at larger
off-shell nucleon momenta. Accordingly the PS coupling is expected to be
important for the two-pion photoproduction, since the far off-shell nucleon is
involved in the intermediate state.

In order to demonstrate the importance of the PS coupling and compare our
theory with the data, we constructed two types of models (PS 
and PV models) by adding the vector meson contributions for the non-resonant
processes (Fig.2) as well as the $\Delta(1232)$ and $N^{\ast}(1520)$
contributions for the resonant processes (Fig.5) to the non-resonant
contributions coming from the $\pi NN$ couplings (Fig.1). Using these models,
we calculated the total cross sections, invariant mass spectra and
helicity-dependent cross sections for various isospin channels. Generally,
these observables are successfully described by the PS model compared with the
PV model except some details. Particularly the calculations of total cross
sections and invariant mass spectra for the $\gamma p\rightarrow\pi^{+}\pi
^{0}n$ and $\gamma n\rightarrow\pi^{-}\pi^{0}p$ channels by the PS model are
in good agreement with the data and on the other hand those by the PV model
are largely underestimated. There are some discrepancies between the
calculations with the PS model and the data in the invariant mass spectra for
the $\gamma p\rightarrow\pi^{0}\pi^{0}p$ channel. Because the magnitude of the
cross section in this channel is very small, the higher order processes might
emerge in the detailed structure. Therefore, we think, it may be necessary to
investigate the mechanism originating with the $\sigma$ meson and the final
state interaction which are not taken into account in our model. Although
there are still unsatisfactory points, we conclude that the PS coupling
certainly plays an important role in the two-pion photoproduction, especially,
the $\gamma p\rightarrow\pi^{+}\pi^{0}n$ and $\gamma n\rightarrow\pi^{-}%
\pi^{0}p$ channels.

Now we would like to make some remarks about our model. First, the difference
between the PS and PV couplings prominently appears in the neutral pion
productions such as $\gamma p\rightarrow\pi^{0}p$ and $\gamma p\rightarrow
\pi^{+}\pi^{0}n$ reactions. This is understood from the fact that it stems
from the anomalous magnetic moment term in the $\gamma NN$ coupling
(Eq.(\ref{gnn})) \cite{kn}. In fact the difference disappears if $F_{2}$ is
set to zero. In our model, $F_{2}$ is taken to be the on-shell value. It would
be interesting to find out the influence of the off-shell effects in the
$\gamma NN$ vertex for the two-pion photoproduction and furthermore to know
how the PS coupling at high photon energies is connected to the PV coupling at
low energies in terms of the off-shellness of the intermediate nucleon. We
note that the off-shell structure of the $\gamma NN$ vertex appears as the
modification of the anomalous magnetic moment term\cite{Kon}.

Secondly, the resonant processes in the present model have been treated in a
naive way. The strong coupling constants for the $N^{\ast}(1520)$ resonance
are determined from its total width and the branching ratios given in Particle
Data Group\cite{PDG}and their form factor ranges are assumed to be consistent
with the nucleon size in quark models. The electromagnetic couplings of the
$N^{\ast}(1520)$ to nucleon are taken from the PDG and their values are real
number. In order to investigate the role of the resonance in detail, the
theory must be refined so as to describe three reactions, i.e., $\pi
N\rightarrow\pi N$, $\gamma N\rightarrow\pi N$ and $\gamma N\rightarrow\pi\pi
N$ , simultaneously. For instance, the effective coupling constant of $\gamma
NN^{\ast}(1520)$ becomes complex due to the interference with the background
process, although its effect is not expected to be large because of the small
imaginary number\cite{tiator}.

Finally, we have not included the $\Delta(1700)$ resonant process in our
model. Nacher et al.\cite{nacher1}have pointed out that the $\Delta(1700)$
resonance has a significant contribution on the $\gamma N\rightarrow\pi\pi N$
reaction cross section due to the strong interference between the $\Delta(1700)$
resonant process and the $\Delta$ Kroll-Ruderman process although the
contribution of only the $\Delta(1700)$ resonant process is almost negligible.
They used the real $\gamma N\Delta(1700)$ coupling constant given in PDG like
the $N^{\ast}(1520)$ resonance in our model. However, it was predicted in the
phenomenological analysis of the $\gamma N\rightarrow\pi N$
reaction\cite{tiator} that its imaginary part was comparable to its real part,
which is quite different from the case of the $\gamma NN^{\ast}(1520)$
coupling. This complex coupling constant, if used, should largely influence
the results of Ref.\cite{nacher1} about the interference effect. Further
investigation is needed to draw a definite conclusion on the $\Delta(1700)$
resonant process.

\begin{acknowledgements}%

We would like to thank V. Metag for providing us the experimental data of the $\gamma
p\rightarrow\pi^{+}%
\pi^{0}n$ reaction.
\end{acknowledgements}


\begin{thebibliography}{9}                                                                                                %

\bibitem {ABB}Aachen-Berlin-Bonn-Hamburg-Heidelberg-M\"{u}nchen Collaboration,
Phys. Rev. \textbf{175}, 1669 (1968).

\bibitem {pia}A.Piazza et al., Nuovo Cim. \textbf{3}, 403\textbf{ }(1970).

\bibitem {gia}G.Gialanella et al., Nuovo Cimento \textbf{63}A, 892 (1969).

\bibitem {car}F.Carbonara et al., Nuovo Cimento \textbf{36}A, 219 (1976).

\bibitem {bra}A. Braghieri et al., Phys. Lett. B\textbf{363}, 46 (1995).

\bibitem {lan}W. Langg\"{a}rtner et al., Phys. Rev. Lett. \textbf{87},
052001-1 (2001).

\bibitem {hae}F. H\"{a}rter et al., Phys. Lett. B\textbf{401}, 229 (1997).

\bibitem {wolf}M. Wolf et al., Eur. Phys. J. A\textbf{9}, 5 (2000).

\bibitem {zab}A.Zabrodin et al., Phys. Rev. C\textbf{55}, R1 (1997).

\bibitem {oset1}J.A.G. Tejedor and E. Oset, Nucl. Phys. A\textbf{571}, 667
(1994); A\textbf{600}, 413 (1997).

\bibitem {laget}L.Y. Murphy and J.M. Laget, Report No. DAPNIA-SPHN-95-42.

\bibitem {ochi1}K. Ochi, M. Hirata, and T. Takaki, Phys. Rev. C\textbf{56},
1472 (1997).

\bibitem {ochi2}M. Hirata, K. Ochi, and T. Takaki, Prog. Theor. Phys.
\textbf{100}, 681 (1998).

\bibitem {nacher1}J.C. Nacher, E. Oset, M.J. Vicente Vacas, and L. Roca, Nucl.
Phys. A\textbf{695}, 295 (2001).

\bibitem {toki}E. Oset, H. Toki, and W. Weise, Phys. Rep. \textbf{83}, 281 (1982).

\bibitem {tiator}D.Drechsel, O.Hanstein, S.S.Kamalov, and L. Tiator, Nucl.
Phys. A\textbf{645}, 145 (1999).

\bibitem {bl}I. Blomqvist and J.M. Laget, Nucl. Phys. A\textbf{280}, 405 (1977).

\bibitem {betz}M. Betz and T.-S.H. Lee, Phys. Rev. C\textbf{23}, 375 (1981).

\bibitem {arima}M.Arima, K.Masutani, and R.Seki, Phys. Rev. C\textbf{51}, 285 (1995).

\bibitem {PDG}Particle Data Group, L.Montanet et al., Phys. Rev. D\textbf{50},
1173 (1994).

\bibitem {ferrari}E.Ferrari and F.Selleri, Nuovo Cimento \textbf{21}, 1028 (1961).

\bibitem {oset-pipi}E.Oset, L. Roca, M.J. Vicente Vacas, and J.C. Nacher, arXiv:nucl-th/0112033.

\bibitem {nacher2}J.C. Nacher and E. Oset, Nucl. Phys. A\textbf{697}, 372 (2002).

\bibitem {lang}M. Lang for the GDH- and A2-Collaboration, \textit{Proc.
GDH2000 Mainz 14-17 June 2000}, \textbf{World Sci.} \textit{ed. Drechsel, Tiator.}

\bibitem {kn}N. Katagiri and T. Takaki (unpublished).

\bibitem {Kon}S. Kondratyuk, G. Martinus, and O. Scholten, Phys. Lett.
B\textbf{418}, 20 (1998).
\end{thebibliography}
\end{document}